\documentclass[aps,prl,reprint,groupedaddress,showpacs,amsmath,amssymb,twocolumn,floatfix]{revtex4-1}
\usepackage{graphicx}
\usepackage{bm}
\usepackage{color}
\usepackage[normalem]{ulem}
\usepackage{amsmath}
\usepackage{tabularx}

\usepackage[colorlinks=true]{hyperref}  
\hypersetup{
    unicode=false,          
    pdftoolbar=true,        
    pdfmenubar=true,        
    pdffitwindow=false,     
    pdfstartview={FitH},    
    pdftitle={XXX},    
    pdfauthor={XXX et al.},     
    pdfsubject={},   
    pdfcreator={},   
    pdfproducer={}, 
    pdfkeywords={} {} {}, 
    pdfnewwindow=true,      
    colorlinks=true,       
    linkcolor=magenta, 
    citecolor=blue,        
    filecolor=magenta,      
    urlcolor=blue           
} 
\usepackage{cleveref}

\newcommand{\Rpara}{$R_{\parallel}$}
\newcommand{\Rperp}{$R_{\perp}$}

\newcommand{\DRR}{$\Delta R / R_0$}

\usepackage{pifont}

\usepackage{bbm}

\newcommand{\Tonset}{$T_{\text{onset}}$}

\begin{document}

\title{Angular Interplay of Nematicity, Superconductivity, and Strange Metallicity in a Moir\'e Flat Band}

\author{Naiyuan J. Zhang$^{1}$}
\author{Pavel A. Nosov$^{2}$}
\author{Ophelia Evelyn Sommer$^{2}$}
\author{Yibang Wang$^{1}$}
\author{K. Watanabe$^{3}$}
\author{T. Taniguchi$^{4}$}
\author{Eslam Khalaf$^{2}$}
\author{J.I.A. Li$^{1}$}
\email{jia\_li@brown.edu}

\affiliation{$^{1}$Department of Physics, Brown University, Providence, RI 02912, USA}
\affiliation{$^{2}$Department of Physics, Harvard University, Cambridge, MA 02138, USA}
\affiliation{$^{3}$Research Center for Functional Materials, National Institute for Materials Science, 1-1 Namiki, Tsukuba 305-0044, Japan}
\affiliation{$^{4}$International Center for Materials Nanoarchitectonics,
National Institute for Materials Science,  1-1 Namiki, Tsukuba 305-0044, Japan}

\date{\today}

\maketitle

\textbf{Superconductivity in strongly correlated electron systems frequently exhibits broken rotational symmetry, raising fundamental questions about the underlying order parameter symmetry. In this work, we demonstrate that electronic nematicity—driven by Coulomb-mediated rotational symmetry breaking—serves as a crucial link to understanding the nature of superconductivity. Utilizing a novel framework of angle-resolved measurement, we reveal an intriguing angular interplay among nematicity, superconductivity, and strange metallicity in magic-angle twisted trilayer graphene. By establishing a direct correlation between the preferred superconducting transport direction and the principal axis of the metallic phase, our findings place strong constraints on the symmetry of the superconducting order parameter. This work introduces a new paradigm for probing the microscopic mechanisms governing  superconductivity in strongly interacting two-dimensional systems.
}

In strongly correlated quantum materials, the interplay between nematicity and superconductivity has been extensively explored as a key pathway to understanding the underlying superconducting  order parameter ~\cite{Kivelson1998nematic,Fradkin2010nematic,Fernandes2014nematicity,Bohmer2022nematicity,Wu2017nematic,Wu2020nematic,Cao2021nematicity}. In high-temperature superconductors ~\cite{Wu2017nematic} and magic-angle twisted bilayer graphene ~\cite{Cao2021nematicity},  anisotropy in superconducting transport is often interpreted as evidence of nodal points in the superconducting energy gap structure. Such observations suggest an unconventional superconducting order parameter that deviates from the isotropic s-wave behavior predicted by conventional Bardeen-Cooper-Schrieffer (BCS) theory.

However, focusing solely on superconducting anisotropy below the transition temperature raises a fundamental question: Can broken rotational symmetry in the superconducting phase serve as a definitive indicator of an unconventional order parameter ~\cite{Wu2017nematic, Cao2021nematicity}? The ambiguity arises when transport anisotropy is already present in the metallic phase above the superconducting transition, suggesting that rotational symmetry breaking occurs independently of superconductivity. The emergence of superconductivity from an anisotropic Fermi sea underscores the necessity of understanding the angular interplay between superconductivity and the metallic phase to fully resolve the origin of superconducting anisotropy.

This scenario raises several outstanding open questions: How does transport anisotropy in the metallic phase influence superconducting transport properties? What constitutes definitive evidence that the superconducting order parameter deviates from the conventional BCS model? How can Coulomb-driven rotational symmetry breaking be distinguished from extrinsic effects such as strain and lattice distortions? Addressing these questions is crucial for gaining deeper insights into the mechanisms governing superconductivity.

Motivated by these outstanding questions, this work investigates the interplay between nematicity and other emergent orders in the moir\'e flat band of magic-angle twisted trilayer graphene (tTLG). In this system, flat energy bands ~\cite{Cao2018a, Bistritzer2011} give rise to a complex interplay between rotational symmetry breaking ~\cite{Jiang2019STM, Rubio2022nematic, Nuckolls2023texture, Cao2021nematicity, Zhang2024nonreciprocity}, strange metallicity ~\cite{Cao2020strange, Jaoui2022linearT, Polshyn2019linearT}, and superconductivity ~\cite{Cao2018b, Yankowitz2019SC, Lu2019SC}.

By characterizing their angular symmetries, we uncover a striking interplay among these intertwined orders. The direction of the most robust superconducting state—characterized by the highest critical temperature and largest critical d.c. supercurrent—aligns with the principal axis of the normal metallic phase associated with the lowest conductivity. Additionally, we demonstrate that the strange metal phenomenon, defined by its characteristic linear-in-T resistivity, is locked to the principal axis of the metallic phase with the highest conductivity. Nematic order, revealed through transport anisotropy, emerges as a unifying thread connecting these phenomena, providing a comprehensive framework for further exploration of their underlying mechanisms.

\begin{figure*}
\includegraphics[width=0.98\linewidth]{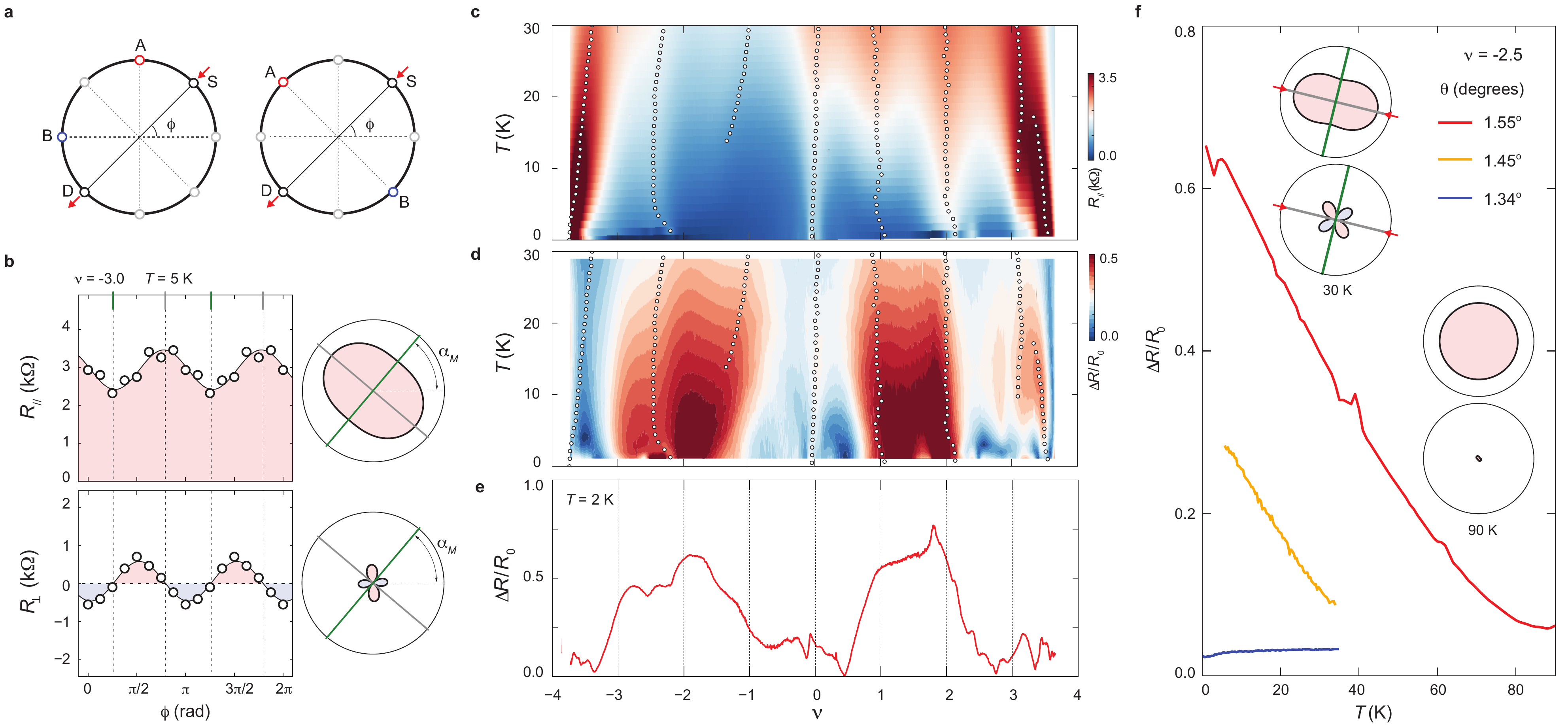}
\caption{\label{fig1}{\bf{Transport anisotropy in the metallic state.}} (a) Schematic of the angle-resolved measurement setup for $V_{\parallel}$ (left) and $V_{\perp}$ (right) in a disk-shaped sample. (b) The angular dependence of linear transport response measured at the moir\'e band filling of $\nu = -3$. Black solid line is the best fit to the angular dependence according to Eq.~1 and ~2.  (c-d) Longitudinal resistance $R_{\parallel}$ (c) and transport anisotropy, defined as the ratio $\Delta R /R_0$ (d), as a function of moir\'e band filling $\nu$ and temperature $T$. Black open circles indicate local maxima in \Rpara, marking boundaries between different isospin ferromagnetic orders ~\cite{Saito2021pomeranchuk,Liu2022DtTLG}. (e) $\Delta R /R_0$ as a function of $\nu$ measured at $T=2$K. (f) $\Delta R /R_0$ as a function of $T$, measured at $\nu=-2.5$ for three samples with different twist angles. All measurements are performed with a small a.c. current bias of $5$ nA. 
}
\end{figure*}

\begin{figure*}
\includegraphics[width=0.87\linewidth]{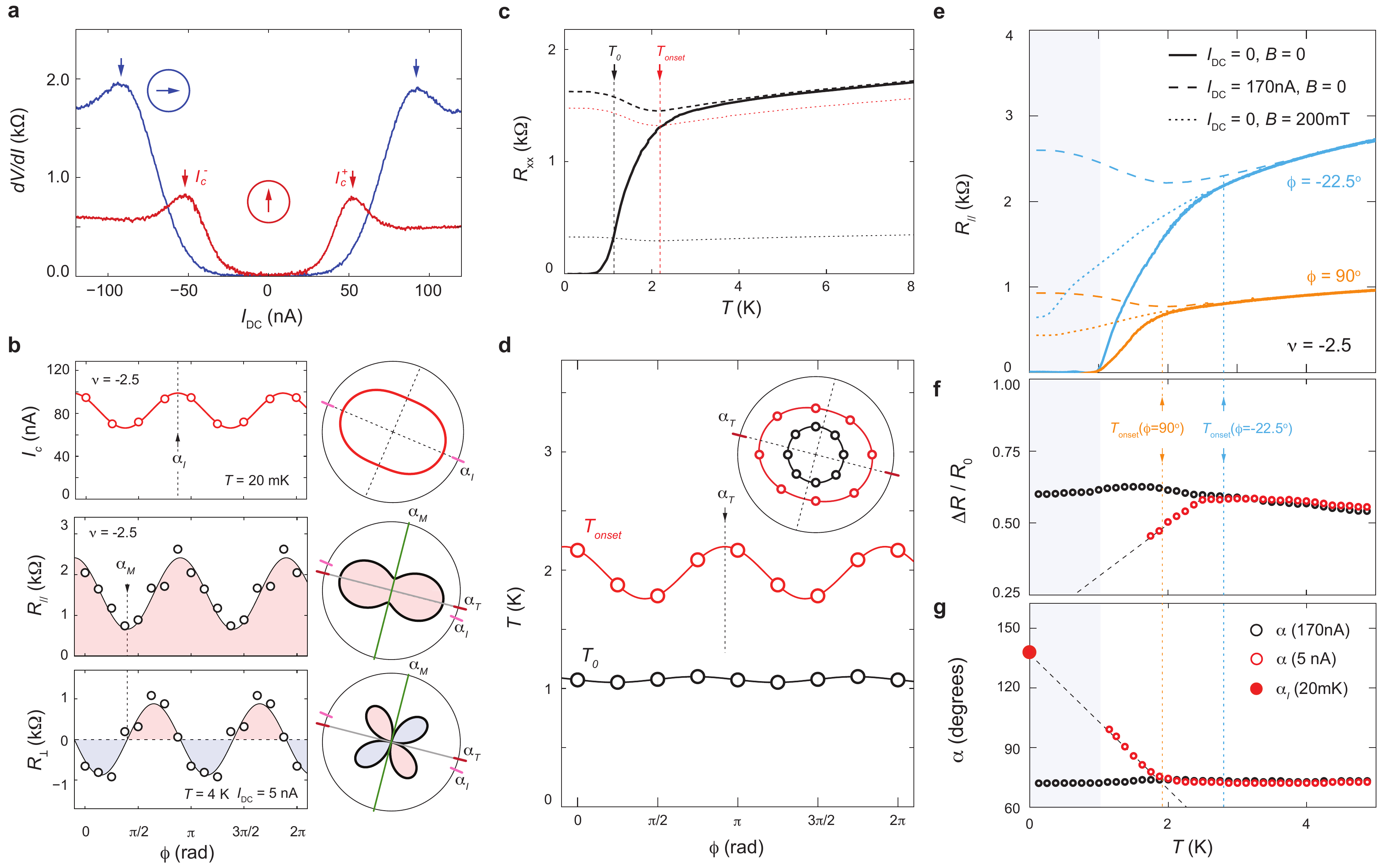}
\caption{\label{fig2}{\bf{Angular interplay between superconducting and metallic phases.}} (a) Differential resistance $dV/dI$ as a function of d.c. current bias $I_{dc}$, measured for different current flow directions. The critical supercurrent $I_c$ is identified as the peak position in $dV/dI$ (indicated by vertical arrows).  (b) Angular dependence of critical supercurrent $I_c$ (top) measured at $T=20$mK, compared to the angular dependence of \Rpara\ (middle) and \Rperp\ (bottom) measured at $T=4$K.  (c) \Rpara\ as a function of temperature (black solid trace) measured with a small a.c. current $I_{ac}=5nA$. The dashed black line represents the temperature dependence of the metallic phase, where superconductivity is suppressed by a large d.c. current, $I_{dc}=170$ nA. The superconducting transition is defined by two temperature points: $T_0$ (where \Rpara\ reaches $10\%$ of the metallic behavior) and $T_{\text{onset}}$ (where it reaches $90\%$), indicated by black and red dotted lines, respectively. (d) Angular dependence of $T_0$ and $T_{\text{onset}}$. Solid traces denote the best fit using $T_0 (\phi)= \bar{T_{0}} + \Delta_T \cos 2 (\phi - \alpha_T)$. (e) \Rpara\ as a function of temperature, measured for different current flow directions. For each current direction, the temperature dependence of the metallic phase response is extracted by suppressing superconductivity using either a magnetic field $B = 200$ mT (dotted lines) or a large d.c. current $I_{dc}=170$ nA (dashed lines). (f) Transport anisotropy strength $\Delta R/R_0$ and (g) principal axis orientation $\alpha$, extracted from the angle-resolved transport data, as a function of temperature. All data are measured at the optimal doping of superconductivity, $\nu=-2.5$.}
\end{figure*}

To perform angle-resolved transport measurements~\cite{Chichinadze2024nonlinearHall,Zhang2024nonreciprocity}, tTLG samples are patterned in a ``sunflower" geometry, featuring a disk-shaped area with eight evenly spaced electrical contacts, as illustrated in Fig.~\ref{fig1}a. In the metallic phase, the conductivity matrix is extracted by fitting the angular oscillation of longitudinal (\Rpara) and transverse (\Rperp), measured as a function of the current flow direction $\phi$ (Fig.~\ref{fig1}b), to the following equations ~\cite{Vafek2023anisotropy}:
\begin{eqnarray}
R_{\parallel}(\phi) &=& R_0 - \Delta R \cos2(\phi-\alpha_M), \\
R_{\perp}(\phi) &=& \Delta R \sin2(\phi-\alpha_M).
\end{eqnarray}
Here, $\Delta R$ represents the oscillation amplitude, $R_0$ is the average value of \Rpara$(\phi)$, and $\alpha_M$ denotes the principal axis orientation of associated with the highest conductivity in the metallic phase. In this context, current flow along $\phi = \alpha_M$ corresponds to the minimum resistivity (or maximum conductivity), whereas flow perpendicular to this axis, at $\phi = \alpha_M + \pi/2$, results in the maximum resistivity (or minimum conductivity). The ratio between maximum and minimum conductivity, qualified by \DRR, offers a direct measure of anisotropy strength. Throughout the discussion, the principal axes corresponding to maximum and minimum conductivity are indicated by green and gray solid lines, respectively, in polar-coordinate plots. Unless otherwise specified, all measurements are performed in the linear response regime using a small current bias of $I < 5$ nA.

At a twist angle of $\theta = 1.55^{\circ}$, close to the expected magic angle of tTLG ~\cite{Park2021tTLG,Hao2021tTLG,Khalaf2019}, we find that the evolution of transport anisotropy is strongly correlated with the  underlying isospin order. The flat energy band of magic-angle moir\'e structures is known to facilitate a cascade of isospin ordered phases. The boundaries between these distinct isospin orders, mapped across the $\nu-T$ map, are delineated by resistance peaks, marked as open black circles in Fig.~\ref{fig1}c-d ~\cite{Saito2021pomeranchuk,Park2021flavour,Rozen2020pomeranchuk,Liu2022DtTLG}.  The most pronounced anisotropy, with \DRR\ $\sim 0.5$, is observed near half-filling of both electron- and hole-doped moir\'e bands (Fig.~\ref{fig1}c), which aligns with an isospin order featuring three-fold degeneracy (see Fig.~\ref{fan}). Conversely, a nearly isotropic transport response is seen in the isospin-unpolarized regime near the charge-neutrality point (CNP) and at the edge of the moir\'e bands (Fig.~\ref{fig1}d). 

Notably, significant anisotropy is observed within the band-filling range $-3 < \nu < -2$, a region associated with a low-temperature superconducting phase. The anisotropy in this regime shows a pronounced dependence on temperature and twist angle, as illustrated in Fig.~\ref{fig1}f. Across three tTLG samples, substantial anisotropy is observed near the magic angle at low temperatures, while it diminishes with increasing temperature or detuning from the magic angle. These findings suggest that transport anisotropy has a Coulomb-driven origin, underscoring the need for further investigations into the interplay between superconductivity and nematicity. To characterize the nature of superconducting anisotropy, we employ two independent methods.

The magnitude of the critical supercurrent, $I_c$, marking the current-driven transition between the superconducting to metallic phases ~\cite{Benyamini2019fragility}, is widely regarded as a direct measure of superconducting stability ~\cite{Cao2018b,Yankowitz2019SC,Lu2019SC,Liu2021DtBLG,Liu2022DtTLG,Lin2022SDE}. In the current-voltage (I-V) curve, $I_c$ corresponds to the peak in differential resistance ($dV/dI$) (Fig.~\ref{fig2}a). Fig.~\ref{fig2}b displays the evolution of $I_c$ with varying current flow direction. Notably, maximum $I_c$, occurring when d.c. current flows along $\phi = \alpha_I$, aligns precisely with the direction of minimum conductivity (maximum resistivity) in the metallic phase.

Taken together, the angular dependence of the  critical current and the normal state conductivity place important constraints on the nature of the superconducting state. First, we note that a simple Ginzburg-Landau analysis which assumes a uniform gap but a directionally-dependent effective mass will predict a critical current that is largest in the direction with largest conductivity, contradicting the observed trend. Accounting for vortices as a mechanism for the critical current will also predict the wrong trend since their transverse motion should be more restricted in the more resistive direction leading to a higher critical current in the \emph{less} resistive direction. 

As demonstrated in the supplemental material, even a pure $d$-wave does not produce anisotropic critical current. Only a $p$-wave or a mixture of $s$ and $d$ (anisotropic $s$) can account for the observed anisotropy. In the latter case, we have to assume a minimal level of anistropy to overcome the anisotropy of the effective mass. We note that the analysis above is based on the assumption that pairing takes place in a single (spinful or spinless) Fermi surface. This is likely the case if the parent normal state spontaneously breaks valley ${\rm U}(1)$ symmetry. In fact, most candidate ground states at low and intermediate strain do indeed involve a degree of intervalley coherence \cite{Bultinck2020, Kwan2021}, which appears on the microscopic scale as a Kekul\'e distortion. Such distortion was recently observed in STM supporting the assumption of intervalley coherence \cite{Nuckolls2023texture}. This analysis is corroborated with the measurement of the superconducting onset temperature, which can be viewed as a direct measure of the gap, as discussed below.

The temperature-driven superconducting transition is characterized by two key temperatures, \Tonset\ and $T_0$. The onset of superconductivity at \Tonset\ is defined by the bifurcation between the $R-T$ curve of the metallic phase (black dashed line in Fig.~\ref{fig2}c) from the superconducting behavior, where \Rpara\ diminishes with lowering temperature (solid black line in Fig.~\ref{fig2}c). Meanwhile, $T_0$ corresponds to the temperature at which \Rpara\ rises above the noise floor. Operationally, \Tonset\ and $T_0$ are defined as the temperatures where \Rpara\ equals $90\%$ and $10\%$ of the metallic phase resistance, respectively.

Fig.~\ref{fig2}d shows the angular dependence of $T_{\text{onset}}$ and $T_0$, extracted from $R-T$ measurements with varying current flow directions. Two key observations emerge from this analysis. First, $T_0$ is nearly isotropic, indicating a uniform sample that supports dissipationless charge transport regardless of the current flow direction. Second, \Tonset\ exhibits a pronounced oscillation as a function of the current flow direction $\phi$. By combining the angular dependence of $T_0$ and \Tonset, we find that the mean-field superconducting transition temperature follows the same angular dependence as \Tonset. As such, the direction of maximum \Tonset, denoted as $\alpha_T$, defines the direction of the most robust superconducting phase.

Remarkably, $\alpha_T$ aligns closely with $\alpha_I$, revealing a deep connection between the mechanism governing the onset of superconducting fluctuation and the low-temperature critical supercurrent. Both $\alpha_T$ and $\alpha_I$ are oriented along the direction of lowest conductivity (or maximum resistivity) in the metallic phase. The fact that the superconducting phase preferentially stabilizes in directions where metallic transport is least efficient introduces an unexpected twist to the angular correlation. This angular interplay between the metallic and superconducting phases points to a shared, underlying symmetry-breaking mechanism, offering profound insights into the intertwined nature of these phases and their fundamental connection to nematic order.

We next demonstrate that the angular interplay between the superconducting and metallic phases gives rise to unique transport properties in the fluctuation regime, which spans the temperature range $T_0 < T < T_{\text{onset}}$ (Fig.~\ref{fig2}e-g). In this regime, the angular dependence of \Rpara\ and \Rperp\ is measured at a fixed temperature, allowing us to extract \DRR\ and $\alpha$ as a function of temperature.  While the angular dependence of the metallic phase, extracted by suppressing superconductivity with a large d.c. current bias ($I_{dc} = 170$ nA, black circles in Fig.~\ref{fig2}f-g), remains largely temperature insensitive, the emergence of superconductivity induces a notable suppression in \DRR, accompanied by a simultaneous rotation in $\alpha$. This provides a direct characterization of the superconducting contribution to angle-dependent transport.

We argue that the evolution of \DRR\ and $\alpha$ in the fluctuation regime provides independent confirmation for the angular interplay between the superconducting and metallic phases. Since the superconducting phase preferentially stabilizes along directions of minimum conductivity in the metallic phase, the onset of superconductivity enhances transport conductivity along this direction by providing a parallel mechanism for transport through superconducting fluctuations (Aslamazov-Larkin effect \cite{Larkin_Varlamov}). Since superconductivity onset is strongest along the most resistive direction in a normal state, that direction would benefit from this parallel channel the most, reducing the disparity between maximum and minimum conductivity. Furthermore, it is also known that in anisotropic superconductors the same fluctuations could also increase transverse resistivity due to scattering of the normal state quasiparticles by the virtual Cooper pairs \cite{Ioffe1993}. Both effects are consistent with observations.

Additionally, because the preferred direction of superconducting transport is rotated by $90^{\circ}$ from the direction of maximum conductivity in the metallic phase, the increasing dominance of superconducting transport in the fluctuation regime naturally shifts the principal axis away from $\alpha_M$ toward $\alpha_I$ and $\alpha_T$. Notably, the $T$-dependence of $\alpha$ in the fluctuation regime extrapolates to $\alpha_I$ and $\alpha_T$ at low temperature, providing additional support for this conjecture.

\begin{figure*}
\includegraphics[width=0.75\linewidth]{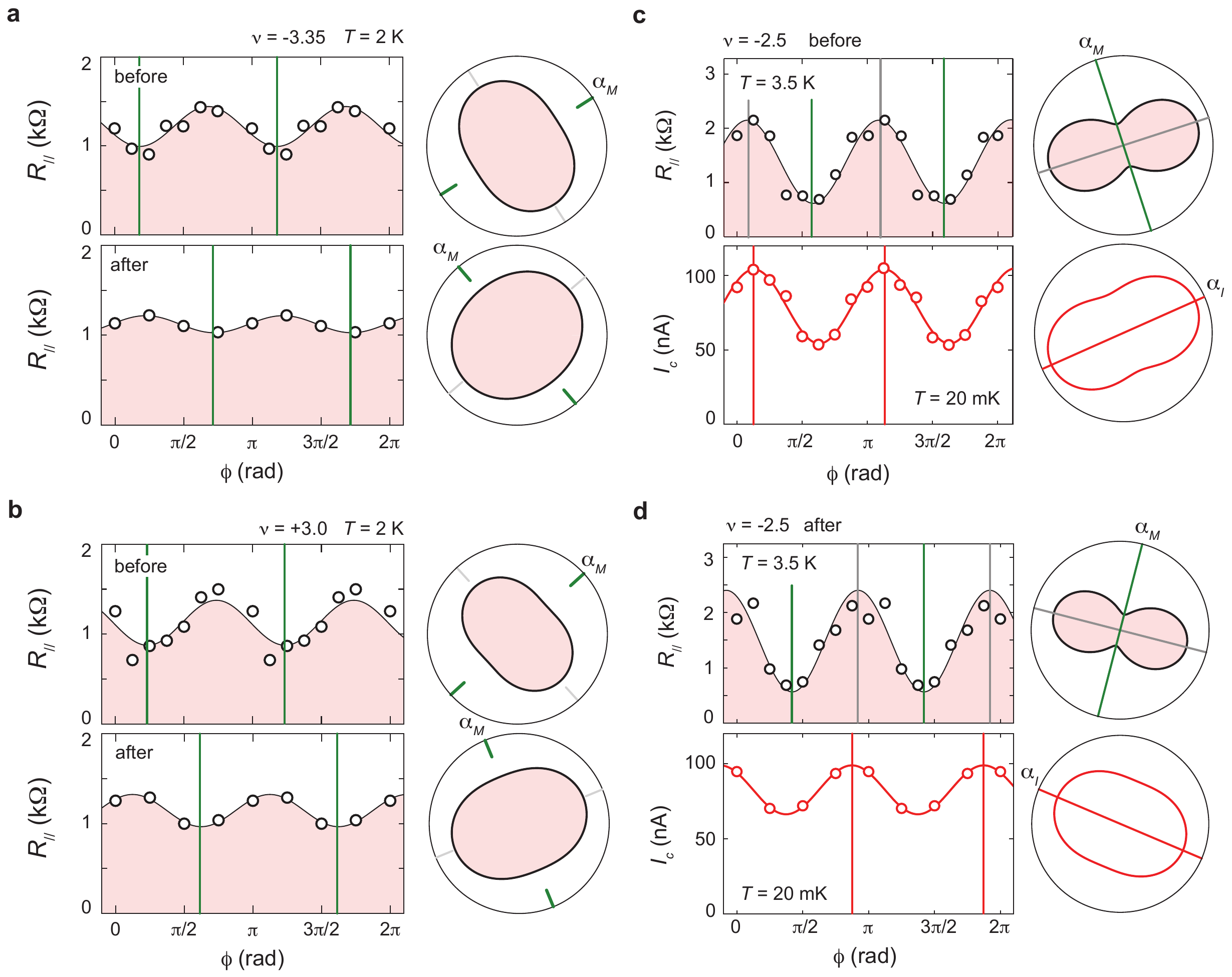}
\caption{\label{fig3}{\bf{Temperature-driven hysteresis in transport anisotropy.}} (a-b) The angular dependence of \Rpara\ before (top) and after (bottom) a thermal cycle up to $T=300$K, measured at $\nu=-3.35$ (a) and $\nu=+3.0$ (b) at $T=2$K. (c-d) The angular dependence of metallic state \Rpara\ at $T=3.5$K (top) and the superconducting state $I_c$ at $T=20$mK (bottom) measured before (c) and after (d) a thermal cycle up to $T=300$K, measured at $\nu=-2.5$. }
\end{figure*}

\begin{figure*}
\includegraphics[width=0.99\linewidth]{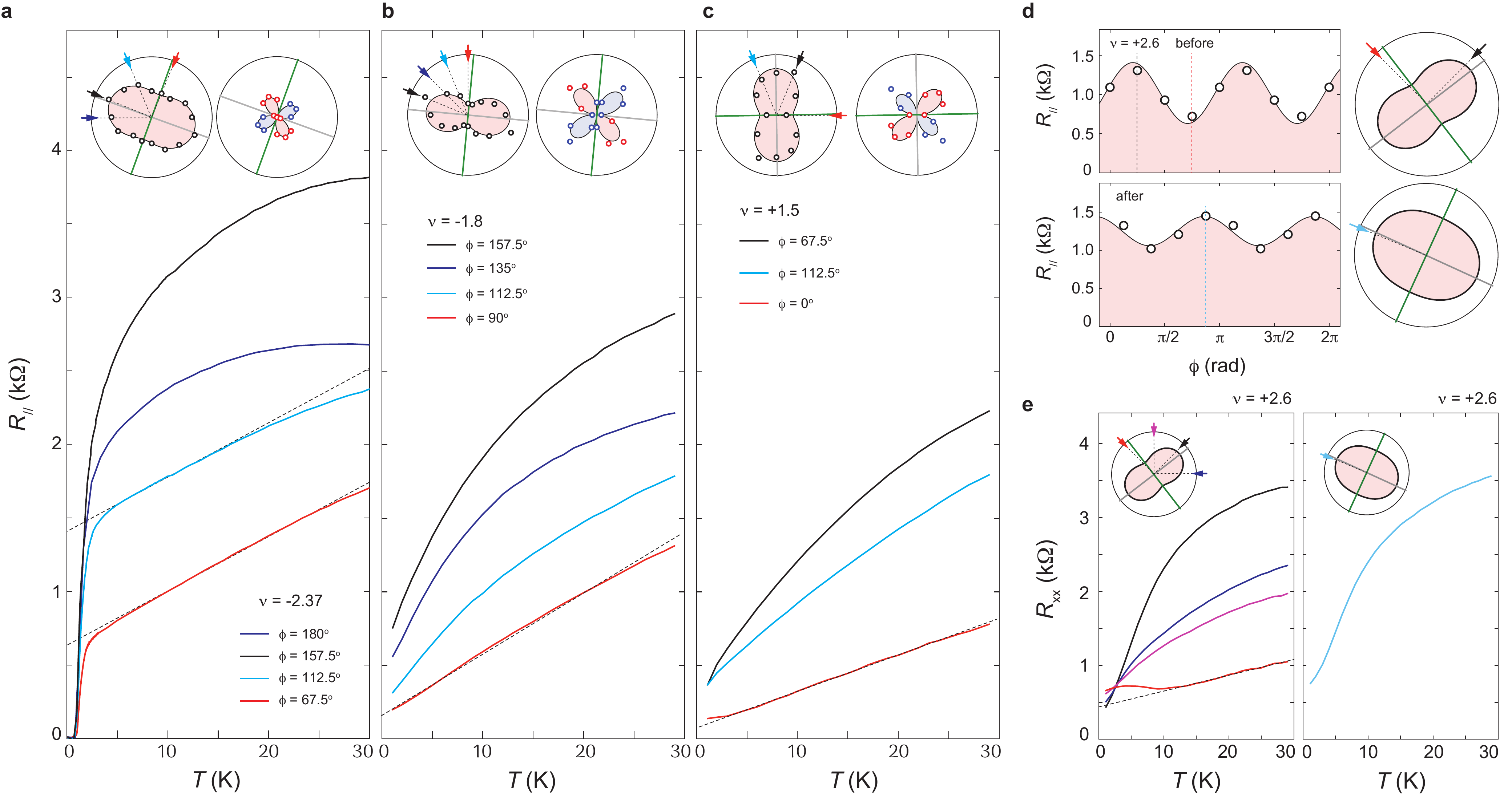}
\caption{\label{fig4}{\bf{Angular interplay of nematicity and strange metallicity. }}  (a-c) \Rpara\ versus $T$ measured with different $\phi$ at (a) $\nu = -2.37$, (b) $\nu = -1.8$, and (c) $\nu = +1.5$. The insets show angle-resolved transport response at $T = 5$ K as the polar-coordinate plots. Green and gray solid lines indicate the principal axes corresponding to maximum and minimum conductivity, respectively. Red arrows mark the direction of maximum conductivity in the metallic phase, which also exhibits the distinctive linear-in-T behavior (red solid lines). (d) Angular dependence of \Rpara\ measured at $\nu = 2.6$ before (top) and after (bottom) temperature-driven hysteresis. (e) $R-T$ curves measured at t $\nu = 2.6$ before (left) and after (right) temperature-driven hysteresis.  }
\end{figure*}

Having established the connection between the angular symmetries of the superconducting and metallic phases, we now address the fundamental question of the origin of transport anisotropy. This question is invariably tied to a compelling puzzle: Is the observed anisotropy a consequence of extrinsic effects, such as strain or lattice distortions, or does it arise intrinsically as a Coulomb-mediated ordering? 

If anisotropy arises from a Coulomb-driven mechanism, the spontaneous breaking of rotational symmetry is expected to produce hysteresis in the transport anisotropy. This hypothesis is supported by the observation of hysteretic transitions in the transport anisotropy, induced by thermal cycles between low temperature and $30$ K. Angle-resolved measurements conducted before and after thermal cycles reveal a notable rotation in the principal axis of the metallic phase, as shown in Fig.~\ref{fig3}a-b (also see Fig.\ref{Metalhysteresis}).  

Notably, the magnitude of anisotropy, characterized by \DRR, also varies after thermal cycling, likely due to the formation of domains with different anisotropic properties. Given the excellent fit with Eq.~1-2, domain formation likely occurs with length scale much smaller than the sample size, preventing the emergence of macroscopic inhomogeneity.

Additionally, temperature-driven hysteresis is observed in the superconducting anisotropy, revealing a striking connection to the hysteretic rotation of the metallic anisotropy. As shown in Fig.~\ref{fig3}d-e, the preferred direction of superconducting transport-defined by the direction of maximum critical supercurrent-closely tracks the rotation of the principal axis in the metallic phase. Both before and after the temperature-driven hysteresis, the maximum critical supercurrent consistently aligns with the current flow direction corresponding to minimum conductivity in the metallic phase. This hysteretic behavior strongly suggests a Coulomb-driven origin underlying the anisotropy observed in the superconducting and metallic phases, rather than being influenced by extrinsic factors such as strain or lattice disorder.


We note that, the observation of superconducting anisotropy alone is insufficient to definitively differentiate between a nodal order parameter and an anisotropic s-wave superconductor. However, the angular interplay between superconductivity and nematicity provides additional, crucial insights that are essential for unraveling the nature of superconductivity.  Within a BCS analysis, mass anisotropy alone is insufficient to produce an anisotropic gap (see Method Section). Given that an isotroptic gap is inconsistent with both the critical current directional dependence and the anisotropy of the onset temperature, we conclude that our data rule out this scenario and strongly suggest that the interaction responsible for superconductivity is strongly anisotropic. This may be a consequence of a strongly anisotropic form factors inherited from the parent phase or because the fluctuations responsible for superconductivity are themselves nematic. 


Beyond its impact on superconducting transport, the anisotropy in the metallic phase gives rise to a distinct angular dependence of strange metal behavior ~\cite{Cao2020strange,Jaoui2022linearT,Polshyn2019linearT}. The linear-in-T behavior consistently appears along the direction of maximum conductivity, whereas the $R-T$ curves exhibit highly nonlinear when the current is misaligned with this direction (Fig.~\ref{fig4}a-c). This correlation persists across a wide range of moir\'e band fillings, despite substantial rotation of the principal axis in transport anisotropy.

The alignment between the linear-in-$T$ behavior and the principal axis of the metallic-phase anisotropy remains robust even in the presence of temperature-driven hysteresis. At $\nu = 2.6$, the direction of maximum conductivity is identified as $\phi = 135^{\circ}$ before the hysteretic transition (top panel of Fig.~\ref{fig4}d). The corresponding $R-T$ curve measured along this direction exhibits a characteristic linear-in-$T$ behavior, as shown in the left panel of Fig.~\ref{fig4}e. However, after the temperature-induced hysteresis, the principal axis of maximum conductivity rotates away from $\phi = 135^{\circ}$ (bottom panel of Fig.~\ref{fig4}d), and the $R-T$ curve near this direction transitions to a highly nonlinear behavior (right panel of Fig.~\ref{fig4}e). 

Notably, in magic-angle graphene moir\'e systems, the investigation of superconductivity has been intricately linked to the ongoing debate surrounding the origin of strange metallicity ~\cite{Cao2020strange,Jaoui2022linearT,Polshyn2019linearT,Sarma2020LinearT,Wu2019LinearT,Sankar2022LinearT,Davis2023LinearT}. Our observations indicate that electronic nematicity serves as a crucial missing link to understanding the relationship between superconductivity and strange metallicity. 

Specifically, along the direction of maximum conductivity, the observation of linear-in-$T$ resistivity coincides with a suppressed superconducting phase, whereas a nonlinear $R-T$ curve and an enhanced superconducting phase are both aligned with the direction of reduced transport efficiency in the metallic phase. 
This misalignment between linear-in-$T$ resistivity and the preferred direction of superconducting transport suggests a potential disconnection between the mechanisms underpinning these two phenomena.

A possible explanation for the observed anisotropy in the linear-in-$T$ resistivity is coexistence of a linear-in-$T$ contribution and other non-linear contributions, with strong directional dependence. In this scenario, the linear-in-$T$ contribution to resistivity is overwhelmed by other contributions in the more resistive direction but not in the least resistive direction. We note that a simple calculation for linear-in-$T$ resistivity using bosonic fluctuations predicts an isotropic scattering rate, even if the form factors for the electron-boson interaction are strongly anisotropic (see SM). This would lead to a  linear-in-$T$ resistivity which is largest in the most resistive direction due to mass anisotropy. To be  consistent with the above explanation, the non-linear contributions to resistivity should have a stronger directional dependence than that suggested by mass anisotropy alone i.e. it suggests an anisotropic scattering rate for the non-linear-in-$T$ contribution.

While our findings do not definitively identify the origin of the strange metal behavior, they underscore the angular degrees of freedom as a previously unrecognized dimension. The angular evolution between linear to nonlinear $R-T$ curves, along with their interplay with nematicity and superconductivity provides critical constraints for future theoretical models aiming to unravel the nature of electronic orders in a moir\'e flat band. 

Beyond nematicity and strange metallicity, the angular degrees of freedom provides new perspectives that could reshape our understanding of 2D superconductivity. For example, the angular oscillation in \Tonset\ exceeds $1$ K, significantly larger than the reported effects typically associated with proximity screening ~\cite{Liu2021DtBLG,Liu2022DtTLG}. This strongly suggests that the angular degrees of freedom should be carefully considered when tuning the stability of the superconducting phase ~\cite{Gao2024screening,Barrier2024screening}. 

Additionally, angle-resolved nonreciprocal measurements reveals an interesting interplay between the zero-field superconducting diode effect and the anisotropy in superconducting transport ~\cite{Lin2022SDE,Zhang2024nonreciprocity}. As shown in Fig.~\ref{Diode}, the direction of maximum nonreciprocity in the critical supercurrent is aligned with the direction of maximum conductivity in the metallic phase, which coincides with the direction of suppressed superconducting transport. Such angular sensitivity highlights the necessity of incorporating directional effects into theoretical and experimental studies of moir\'e flat-band physics.

\section*{acknowledgments}
This material is based on the work supported by the Air Force Office of Scientific Research under award no. FA9550-23-1-0482. N.J.Z., and J.I.A.L. acknowledge support from the Air Force Office of Scientific Research. Y.W. and J.I.A.L. acknowledges partial support from the National Science Foundation under NSF Award NSF DMR-2143384.  K.W. and T.T. acknowledge support from the JSPS KAKENHI (Grant Numbers 21H05233 and 23H02052) and World Premier International Research Center Initiative (WPI), MEXT, Japan. Part of this work was enabled by the use of pyscan (github.com/sandialabs/pyscan), scientific measurement software made available by the Center for Integrated Nanotechnologies, an Office of Science User Facility operated for the U.S. Department of Energy.

\bibliography{Li_ref}

\newpage
\clearpage

\section*{Method}

\renewcommand{\thefigure}{M\arabic{figure}}
\def\theequation{M\arabic{equation}} 
\def\thetable{M\Roman{table}}
\setcounter{figure}{0}
\setcounter{equation}{0}

In this section, we provide detailed discussions to further substantiate results reported in the main text. This section offers a comprehensive review, summarizing the notations employed and elaborating on the angle-resolved transport response. More specifically, we provide further details on the evolution of transport anisotropy with varying carrier density and temperature. We also expand on the measurement procedures associated with the temperature-driven hysteresis.  For additional information and in-depth analysis, readers are directed to the Supplementary Materials ~\cite{SI}.

\subsection{I. Theoretical Analysis}
In this section we present complementary theoretical considerations that address the interplay between effective mass anisotropy, pairing symmetry, and transport in anisotropic superconductors. Our starting assumption is that the dominant contributions to transport and pairing comes from a single (possibly spin degenerate) Fermi surface. It is possible to generalize our discussion to two concentric Fermi surfaces provided that we assume pairing is purely intraband and the contributions of the two Fermi surfaces simply add. Importantly, this does \emph{not} include the possibility of intervalley pairing on top of a valley symmetric state. This is supported by experimental evidence from STM of microscopic Kekul\'e distortion  \cite{Nuckolls2023texture} which implies intervalley coherence in the normal state. Intervalley coherence breaks valley ${\rm U}(1)$ symmetry mixing the Fermi surfaces of the two valleys into a single spin degenerate Fermi surface or two concentric spin-polarized Fermi surfaces. We note that intervalley coherences is consistent with most candidate ground states discussed in literature \cite{Bultinck2020, Kwan2021}. We assume that the anisotropy in normal-state resistance primarily arises from effective mass anisotropy (as expected for a nematic state) and analyze the constraints that experimental observations impose on the structure of the pairing interactions within this framework. Using elementary arguments, we show that: (i) within BCS theory, anisotropy in quasiparticle dispersion does not induce gap anisotropy if the pairing form factors are isotropic, (ii) the anisotropy in the critical supercurrent calculated within the Ginzburg–Landau theory is consistent with the experimentally observed behavior provided that the gap symmetry is either an extended s-wave (``s+d") or a p-wave (both of which are expected from pairing mediated by nematic fluctuations \cite{Lederer2015}), and (iii) the linear-in-$T$ scattering rate from classical bosonic fluctuations with nematic form-factors is largely isotropic.

The first point (i) follows trivially from the linearized BCS gap equation, 
\begin{equation}\label{eq:BCS}
    \Delta_{\bm{k}} = -\sum\limits_{\bm{k}'} \lambda_{\bm{k},\bm{k}'}\frac{\Delta_{\bm{k}'}}{2E_{\bm{k}'}}\tanh \frac{E_{\bm{k}'}}{2T},
\end{equation}
where $E_{\bm{k}}=\sqrt{\xi_{\bm{k}}^2+|\Delta_{\bm{k}}|^2}$ is the quasiparticle energy, $\xi_{\bm{k}}= k_x^2/2m_x +k_y^2/2m_y -E_F$ is the dispersion relation measured from Fermi energy $E_F$, and $\lambda_{\bm{k},\bm{k}'}$ is the pairing interaction. If $\lambda_{\bm{k},\bm{k}'}$ is independent of the position on the Fermi surface (i.e. $\lambda_{\bm{k},\bm{k}'}\equiv \lambda$), then the anisotropic contributions on the right-hand side of Eq.~\eqref{eq:BCS} average out, yielding a unique, isotropic gap function $\Delta_{\bm{k}}\equiv \Delta$.

Next, we turn to the second point (ii) and discuss the critical supercurrent anisotropy. Following \cite{Gorkov1964,Kogan2002}, we employ the effective Landau-Ginzburg theory which is valid when the condensate coherence length $\xi$ is larger than the Fermi wavelength $\lambda_F$, and when the condensate is weak. Let $\Delta_{\bm{k}}(\bm{r})$ be the local pairing strength at position $\bm{r}$, which we further decompose as $\Delta_{\bm{k}}(\bm{r})= \Omega_{\bm{k}}\Delta(\bm{r})$, with $\Omega_{\bm{k}}$ describing the relative momentum structure of the gap function, and $\Delta(\bm{r})$ encodes smooth variations of the gap phase and magnitude on scales longer than the coherence length. The effective theory for the latter has free energy 
\begin{align}
    F[\Delta]=\int d^2\bm{r} \left[\alpha|\Delta|^2+\frac{\beta}{2}|\Delta|^4+\frac{1}{2}K^{\mu\nu}\nabla_\mu\bar{\Delta}\nabla_\nu\Delta\right]\;,
\end{align}
where $\alpha$ and $\beta$ depend on the microscopic details (e.g. presence of disorder). Upon coupling to a vector potential $\bm{A}$, the derivatives shift by minimal substitution $\nabla_\mu\mapsto \nabla_\mu \mp2ie  A_\mu$. Notably, the stiffness $K^{\mu\nu}$ is allowed to be \emph{anisotropic}, due to anisotropy in both the single-particle dispersion $\xi_{\bm{k}}$ and the gap $\Omega_{\bm{k}}$. Close to the transition, $K^{\mu\mu}$ can be expressed as \cite{Kogan2002}
\begin{equation}\label{eq:Kmumu}
  K^{\mu\mu}\propto \langle|\Omega_{\bm{k}}|^2 v_F^\mu v_F^\mu\rangle_{FS} =\int \frac{d^2k}{(2\pi)^2} \delta(\xi_{\bm{k}}) |\Omega_{\bm{k}}|^2  \frac{k_\mu^2}{m^2_\mu}\;,
\end{equation}
where $v_F^\mu{=}\partial_{k_\mu}\xi_{\bm{k}}{=}k_\mu /m_\mu$. Let $\Delta=|\Delta|e^{i\phi}$. The supercurrent in the absence of an external field is related to the gap via the equation 
\begin{equation}\label{eq:current}
    j^\mu=\frac{\delta F}{\delta A_\mu}=2 e|\Delta|^2K^{\mu\nu}\nabla_\nu\phi\;.
\end{equation}
To deduce the relation between the stiffness $K^{\mu\nu}$ and the critical current, we follow the standard procedure and  minimize the free energy $F$ on a square geometry of side lengths $L$ that aligns with the principal directions of $K^{\mu\nu}$ which we will denote $x,y$. We impose boundary conditions that  $\nabla_y\phi|_{y=0,L}=0$ while $\phi|_{x=0}=0$ and $\phi|_{x=L}=k_\phi L$. This can be achieved by a constant phase gradient, which leaves the bulk homogeneous, having gap magnitude $|\Delta|$. After minimizing the free energy the gap becomes  $|\Delta|=\sqrt{(|\alpha|-K^{xx}k_\phi^2/2)/\beta}$. Using Eq.~\eqref{eq:current} we can deduce the supercurrent as a function of $k_\phi$. Its maximal value then becomes $ j_c^x{=}\frac{4 e|\alpha|}{3\beta}\sqrt{\frac{2|\alpha|K^{xx}}{3}}{\propto}  \sqrt{K^{xx}}$, and similarly along the $y$ direction. The ratio of the critical supercurrents in two principal directions then becomes
\begin{equation}\label{eq:current_max}
    \frac{j_c^x}{j_c^y} =   \sqrt{\frac{K^{xx}}{K^{yy}}}\;.
\end{equation}
 Crucially this means that in the regime where Landau-Ginzburg theory is valid, the critical supercurrent will be largest in the principal axis where the stiffness $K$ is largest.

From Eq.\eqref{eq:Kmumu}, we observe that in the absence of any gap anisotropy (i.e., when $\Omega_{\bm k}\equiv 1$), the stiffness $K$ is maximized along the direction of highest mass. Indeed, we have $K^{xx}\propto\langle v_F^xv_F^x\rangle_{FS} \propto 1/\sqrt{\eta} $ and $ K^{yy}\propto \langle v_F^yv_F^y\rangle_{FS} \propto \sqrt{\eta} $, where  $\eta\equiv m_x/m_y$ is the mass anisotropy parameter. On the other hand, assuming that the transport data in the normal state can be explained by effective mass anisotropy and the Drude theory, we also find the resistivity ratio as
\begin{equation}\label{eq:Drude}
    \rho_{\mu\mu} = \frac{m_{\mu}}{ne^2\tau},\quad \rightarrow \quad \frac{\rho_{xx}}{\rho_{yy}}=\frac{m_x}{m_y}\equiv \eta\;,
\end{equation}
where $n$ is the density and $\tau$ is the scattering time. Thus, in the isotropic gap case, we find
\begin{equation}\label{eq:iso}
    \left.\frac{j_c^x}{j_c^y}\right|_{\rm s-wave} = \frac{1}{\sqrt{\eta}}=   \sqrt{\frac{\rho_y}{\rho_x}}\;.
\end{equation}
Therefore, the critical supercurrent is highest in the more conductive direction. This result is opposite to the observed trend.

It is thus natural to ask what gap symmetries are compatible with the fact that the highest critical current is in the more resistive direction. From Eq.~\eqref{eq:Kmumu}, we obtain 
\begin{equation}
   \begin{aligned}
        \langle|\Omega_{\bm{k}}|^2 v_F^xv_F^x\rangle_{FS} &\propto \int\limits_0^{2\pi} d\theta \frac{|\Omega(\theta)|^2 \cos^2\theta}{\left(\cos^2\theta +\eta \sin^2\theta\right)^2}\;,\\
        \langle|\Omega_{\bm{k}}|^2 v_F^yv_F^y\rangle_{FS} &\propto  \eta^2 \int\limits_0^{2\pi}  d\theta \frac{|\Omega(\theta)|^2 \sin^2\theta}{\left(\sin^2\theta +\eta \sin^2\theta\right)^2}\;,
   \end{aligned} 
\end{equation}
where $\theta$ is the angle on the Fermi surface.  First, we consider the $l=2$ angular momentum component $\Omega(\theta)=\cos(2\theta)$ corresponding to a d-wave symmetry. After performing a simple integration and using Eq.~\eqref{eq:current_max}, we find
\begin{equation}\label{eq:j_c_dwave}
\left.\frac{j^{x}_c}{j^{y}_c}\right|_{\rm d-wave} = \sqrt{\frac{\eta ^{3/2}+3 \eta -\sqrt{\eta }+1}{\eta  \left(\eta ^{3/2}+3 \sqrt{\eta }-\eta +1\right)}}\;.
\end{equation}
Despite a more complicated dependence on $\eta$, the behavior is actually very similar to the isotropic case: the critical current is still smallest in the most resistive direction. We also note that adding a phase, i.e. $\Omega(\theta)=\cos(2\theta+\phi)$, does not change this conclusion. Repeating this exercise for higher order harmonics ($l=3,4,...$), we find that all of them lead to the same behavior. This can already be seen in the limit of small mass anisotropy by expanding Eq.~\eqref{eq:current_max} around $\eta\approx 1$:
\begin{equation}\label{eq:small_eta}
   \frac{j^{x}_c}{j^{y}_c} \approx \sqrt{\frac{J_1}{J_0-J_1}}\Big[1-\frac{J_0 (J_0-J_2)}{4 J_1(J_0-J_1)} (\eta-1)\Big]+\mathcal{O}((\eta-1)^2)
\end{equation}
where $J_s=\int d\theta |\Omega(\theta)|^2\cos^2 (s\theta)$. For $\Omega(\theta)=\cos(l\theta+\phi)$ and $l\geq 2$, we find $J_0=\pi$, $J_1=\pi/2$, $J_2=\pi/2+(\pi/4)\delta_{l,2} \cos(2\phi)$. From Eq.~\eqref{eq:small_eta} we thus find
\begin{equation}
   \left.\frac{j^{x}_c}{j^{y}_c}\right|_{l=2,3,...} \approx 1-\frac{1}{2}\left(1-\frac{\delta_{l,2}}{2}\cos2\phi\right)(\eta-1)\;.
\end{equation}
The coefficient in front of the $(\eta-1)$ term is always negative. Therefore, there is no range of mass anisotropy such that the critical supercurrent is the highest in the least conducting direction. The only pure harmonic that produces the desired trend is $p$-wave, i.e. $l=1$. In this case, we find $J_0=\pi$, $J_1=\pi/2{+}(\pi/4) \cos(2\phi)$, and $J_2=\pi/2$. Form Eq.~\eqref{eq:small_eta} we obtain
\begin{equation}
   \left.\frac{j^{x}_c}{j^{y}_c}\right|_{\rm p-wave} \approx \sqrt{\frac{2+\cos 2\phi }{2-\cos 2\phi}}\left[1-\frac{2(\eta-1)}{\left(4-\cos^2(2\phi)\right)}\right]
\end{equation}
From the first term in this expression we see that the critical current is anisotropic even for $\eta=1$ (i.e. in the absence of any mass anisotropy), unless $\phi=\pi/4$. Another option that could in principle be consistent with the observed behavior is a mixture of the $l=0$ and $l=2$ components $\Omega(\theta)=1+\beta \cos(2\theta+\phi)$ (i.e. extended s-wave). In this case, the leading term in Eq.~\eqref{eq:small_eta} is $\sqrt{(2+\beta^2+2\beta\cos\phi)/(2+\beta^2-2\beta\cos\phi)}$, which by itself implies  current anisotropy unless $\phi$ is $\pi/2$.

Finally, let us comment on our point (iii) concerning the momentum-dependence of the scattering rate in the regime when the electrons are interacting with classical bosonic fluctuations of the nematic order parameter. At the lowest order in perturbation theory, this rate can be estimated from the imaginary part of the electron self-energy
\begin{equation}\label{eq:Sigma}
    \frac{1}{\tau_{\bm{k}}(T)}\propto -T \int \frac{d^2p}{(2\pi)^2}f^2_{(\bm{k}-\bm{p})/2}\operatorname{Im}G_{\bm{p}}^R(\omega{=}0),
\end{equation}
where $f_{\bm{k}} = (k_x^2-k_y^2)/(k_x^2+k_y^2)$ encodes the nematic form factor, and $G^R_{\bm{p}}(\omega{=}0)$ is the retarded Green's function for electrons. In Eq.~\eqref{eq:Sigma} we already assumed that the bosonic spectral function is essentially featureless because the typical energy transfer $ \omega\sim T \gg \omega_N = \max_{\bm{k}}\{\omega_{\bm{k}}\}$, where $\omega_N$ is some characteristic frequency of the nematic fluctuations, and $\omega_{\bm{k}}$ is their dispersion relation. If the system is sufficiently clean, then $\operatorname{Im}G^R_{\bm{p}}(\omega{=}0)=-\pi\delta(\xi_{\bm{p}}) $, and thus the scattering rate is proportional to the average $\langle f^2_{(\bm{k}-\bm{p})/2}\rangle_{FS}$ over all momenta $\bm{p}$ on the Fermi surface. In the limit of small mass anisotropy, $\eta\approx 1$, and when $\bm{k}$ is on the Fermi surface, we find $\langle f^2_{(\bm{k}-\bm{p})/2}\rangle_{ FS} \approx \langle \cos^2(\theta_{\bm{k}}+\theta_{\bm{p}})\rangle_{ FS}$ which is $\bm{k}$-independent. For finite $\eta$, this average yields $\langle f^2_{(\bm{k}-\bm{p})/2}\rangle_{FS}\propto \frac{  (\eta +1)}{\left(1+2\sqrt{\eta }+\eta\right) \sqrt{\eta }}$, i.e. again we obtain a $\bm{k}$-independent rate $\tau^{-1}(T)\sim T$. This behavior is very different from the low-temperature regime, $T\ll \omega_N$, where the nematic fluctuations should be treated as dynamical degrees of freedom. In this case, the self-energy is strongly anisotropic \cite{Oganesyan2001}, but it has a different temperature dependence. Using the Drude formula Eq.~\eqref{eq:Drude}, we then conclude that the angular dependence of the prefactor in front of the $T-$linear term would mostly follow the behavior of the effective mass.

\subsection{II. Transport anisotropy in the metallic phase}

\begin{figure*}
\includegraphics[width=0.87\linewidth]{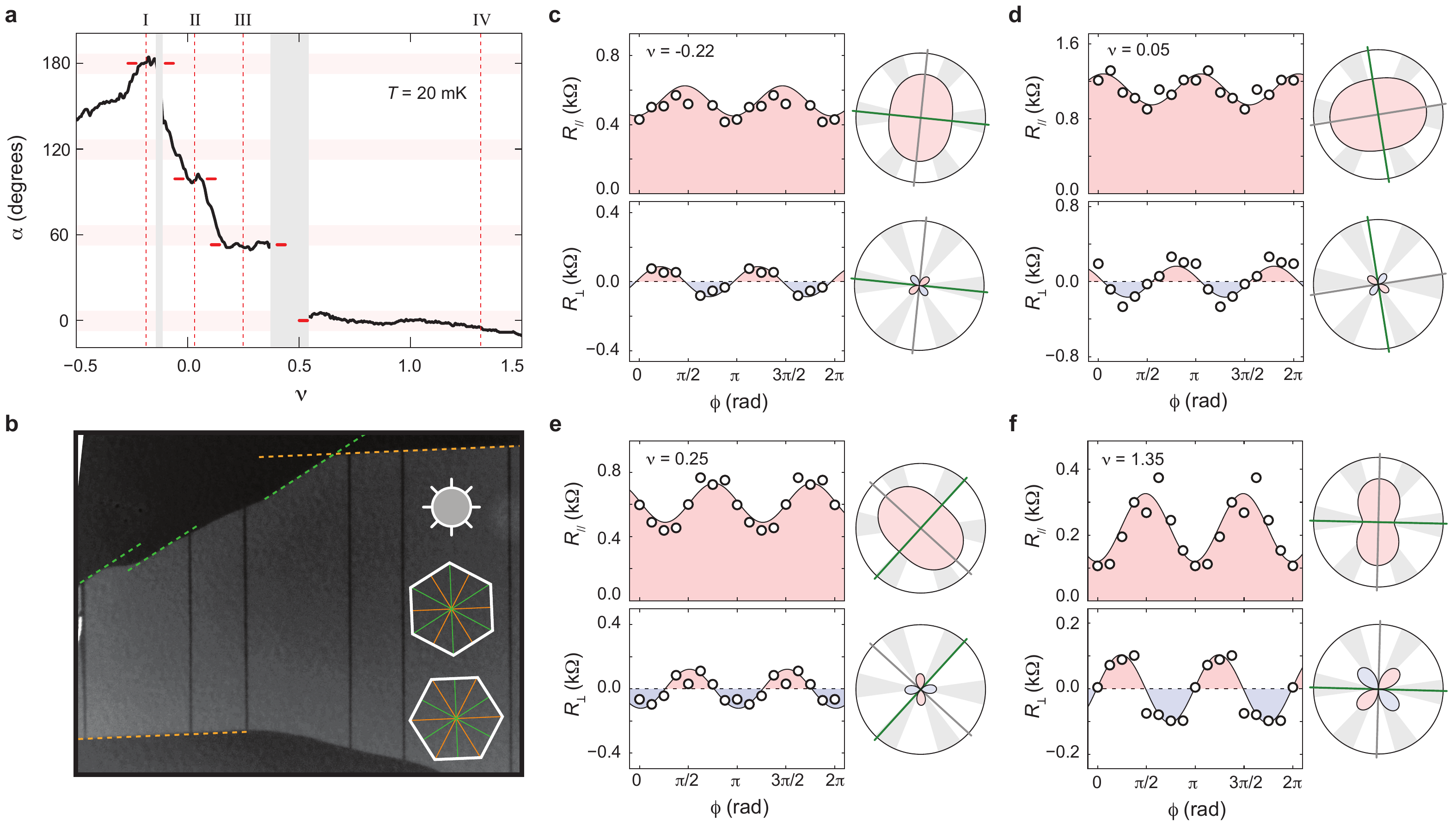}
\caption{\label{Axis}{\bf{Transport anisotropy and crystallographic orientations.}}  (a) The principal axis of transport anisotropy, $\alpha$, as a function of moir\'e filling $\nu$. Horizontal pink stripes mark the orientations of the lattice axes determined from panel (b). (b) Optical image of the graphene crystal. The orientations of the lattice axes are indicated by long straight edges of the crystal, marked by orange and green dashed lines. (c-f) Angular dependence of \Rpara\ and \Rperp\ measured at different $\nu$, exhibiting rotations in the principal axis of anisotropy.  }
\end{figure*}

\begin{figure*}
\includegraphics[width=0.74\linewidth]{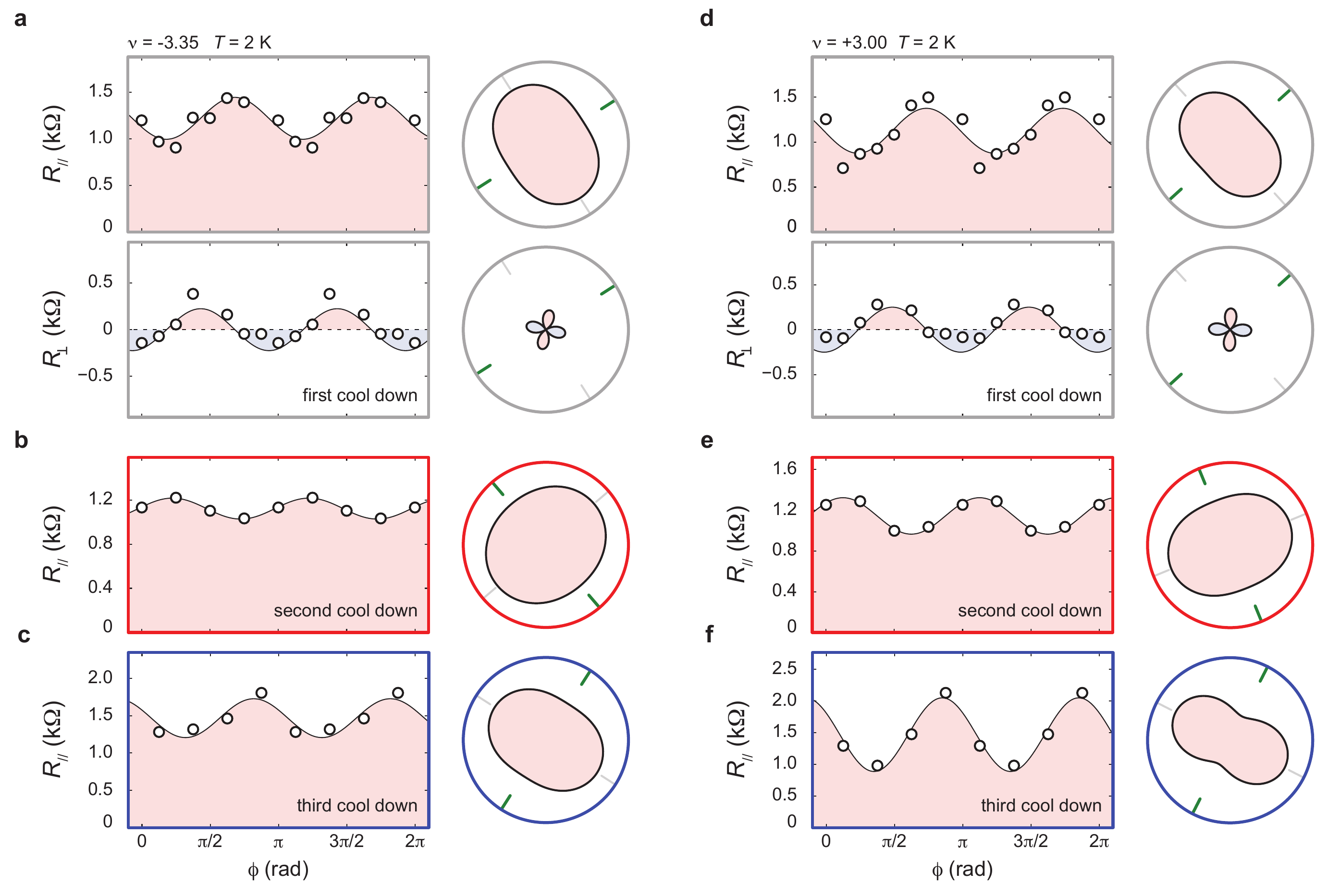}
\caption{\label{Metalhysteresis}{\bf{Temperature-driven hysteresis in the metallic phase.}} The angular dependence of transport responses measured at (a-c) $\nu=-3.35$ and (d-f) $\nu=+3.00$ after consecutive temperature cycles. During each cycle, the temperature is raised to $30$ K and subsequently cooled back down to $T = 2$ K. Angle-resolved transport measurements are performed after the sample reaches $T = 2$ K following each cycle.  The black solid traces denote the best fit of the angular dependence according to Eq.~1 and ~2. The orientation of the principal axis is highlighted by green solid lines in the polar-coordinate plots. }
\end{figure*}

Here, we examine the potential alignment between the principal axis of the metallic phase anisotropy and high-symmetry directions of the moir\'e superlattice. 

As shown in Fig.~\ref{Axis}a, $\alpha$ undergoes a cascade of discrete rotations with varying carrier density $\nu$, resulting in several plateaus indicated by horizontal red solid lines in Fig.~\ref{Axis}a. The angular-dependent transport response for each plateau is presented in Fig.~\ref{Axis}c-f, where solid black lines represent the best fit using Eq.~1-2. The orientation of the principal axis, defined as the direction of maximum conductivity in the metallic phase, is marked by the solid green lines in the polar-coordinate plots. Notably, the principal axis rotates by approximately $60^{\circ}$ between adjacent plateaus. This is reminiscent of previous observations in angle-resolved nonlinear transport measurements in multilayer graphene \cite{Zhang2024nonreciprocity,Lin2023momentum,Chichinadze2024nonlinearHall}.

Fig.~\ref{Axis}b displays the optical image of the monolayer graphene used for fabrication, where the straight edges, marked by orange and green dashed lines, indicate the high-symmetry axes of the the graphene lattice. These edges remain clearly visible after the stacking process, defining the orientation of the moir\'e superlattice. The upper inset of Fig.~\ref{Axis}b highlights the alignment of the sunflower sample geometry relative to the crystallographic axis of the graphene moir\'e structure. Based on this alignment, the high-symmetry directions of the moir\'e superlattice are marked by  horizontal stripes in Fig.~\ref{Axis}b. 

Given the experimental resolution in angle-resolved transport measurements, which arises from the finite width of each graphene contact ~\cite{Chichinadze2024nonlinearHall}, combined with the uncertainty in extracting $\alpha$ from the angle-resolved transport response, we use gray-shaded cones in the polar plots (Figs.~\ref{Axis}c-f) to represent the angular uncertainty in determining the principal axis orientation. Within this context, the plateaus in $\alpha$ show a strong alignment with these high-symmetry directions of moir\'e superlattice, indicating an intrinsic connection between the electronic anisotropy and the underlying lattice.

\begin{figure*}
\includegraphics[width=0.65\linewidth]{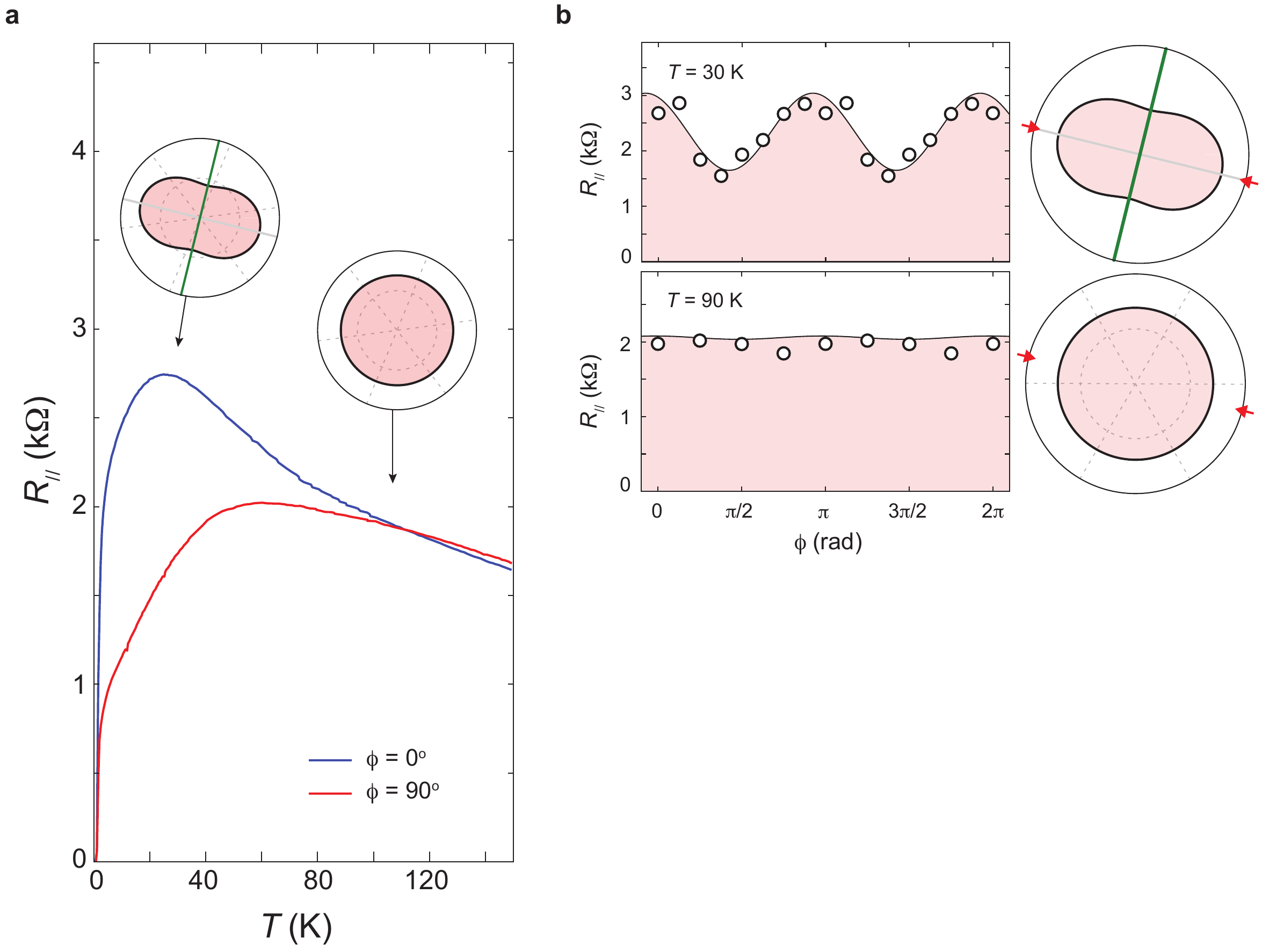}
\caption{\label{FullT}{\bf{$R-T$ curves over a larger temperature window.}} (a-b) The temperature dependence of (a) \DRR\ and (b) $\alpha$ as a function $T$ up to $200$ K. This measurment is performed at the optimal superconducting doping of $\nu = -2.5$. Insets: the angular dependence of transport response in the metallic phase measured at $30$ K, $90$ K, and $150$ K. (c-e) Angular dependence of \Rpara\ and \Rperp\ measured at $T=30$K (c), $T=90$K (d), and $T=150$K (e).  }
\end{figure*}



Fig.\ref{Metalhysteresis} displays multiple temperature-driven transitions, characterized by rotations in the principal axis of the associated transport anisotropy, observed at different moir\'e band fillings. 

These transitions are uncovered through angle-resolved transport measurements conducted over three consecutive temperature cycles. During each cycle, the temperature is raised to $30$ K and subsequently cooled back down to $T = 2$ K. Angle-resolved transport measurements are performed after the sample reaches $T = 2$ K following each cycle, as illustrated in Fig.\ref{Metalhysteresis}. At each band filling, the principal axis undergoes prominent rotation between temperature cycles.  

Apart from the rotation of the principal axis, the strength of anisotropy, characterized by \DRR, also exhibits notable changes before and after the temperature cycles. For example, at $\nu = -3.35$, \DRR\  decreases from $0.18$ in the first cool down (Fig.~\ref{Metalhysteresis}a) to $0.085$ in the second cool down (Fig.~\ref{Metalhysteresis}b).

A plausible explanation for the rotations in $\alpha$, observed as a function of varying $\nu$ or through hysteretic transitions, involves Coulomb-driven instabilities in momentum space ~\cite{Chichinadze2024nonlinearHall,Dong2021momentum,Jung2015momentum,Huang2023momentum}. Strong Coulomb interactions can distort the underlying Fermi surface, causing charge carriers to preferentially accumulate along high-symmetry directions in the Brillouin zone. These distortion generates transport anisotropy, with the direction of maximum conductivity aligning with high-symmetry axes of the moir\'e superlattice. 

Importantly, such Coulomb-induced  Fermi surface distortions are challenging to detect using conventional techniques like quantum oscillation measurements. In contrast, angle-resolved transport measurements offer superior sensitivity to these subtle distortions, providing a powerful tool to probe the resulting anisotropic behavior. Recently, investigations of angle-resolved transport responses in the nonlinear regime have sparked discussions about Coulomb-driven Fermi surface distortions as a possible origin of the nonlinear Hall effect \cite{Chichinadze2024nonlinearHall}. Our findings in the linear transport regime provide experimental indications that align with this perspective, further supporting the role of Coulomb interactions in shaping anisotropic transport behavior.

Fig.~\ref{FullT} shows $R-T$ curves measured along different current flow directions over a larger temperature window. Around $T = 90$ K, the bifurcation between two $R-T$ curves indicates the onset of transport anisotropy. While angle-resolved transport measurement at $T = 30$ K reveals a highly anisotropic response (top panel of Fig.~\ref{FullT}b), a mostly isotropic transport response is observed at $T = 90$ K.

\subsection{III. Across the superconducting regime}

\begin{figure*}
\includegraphics[width=0.95\linewidth]{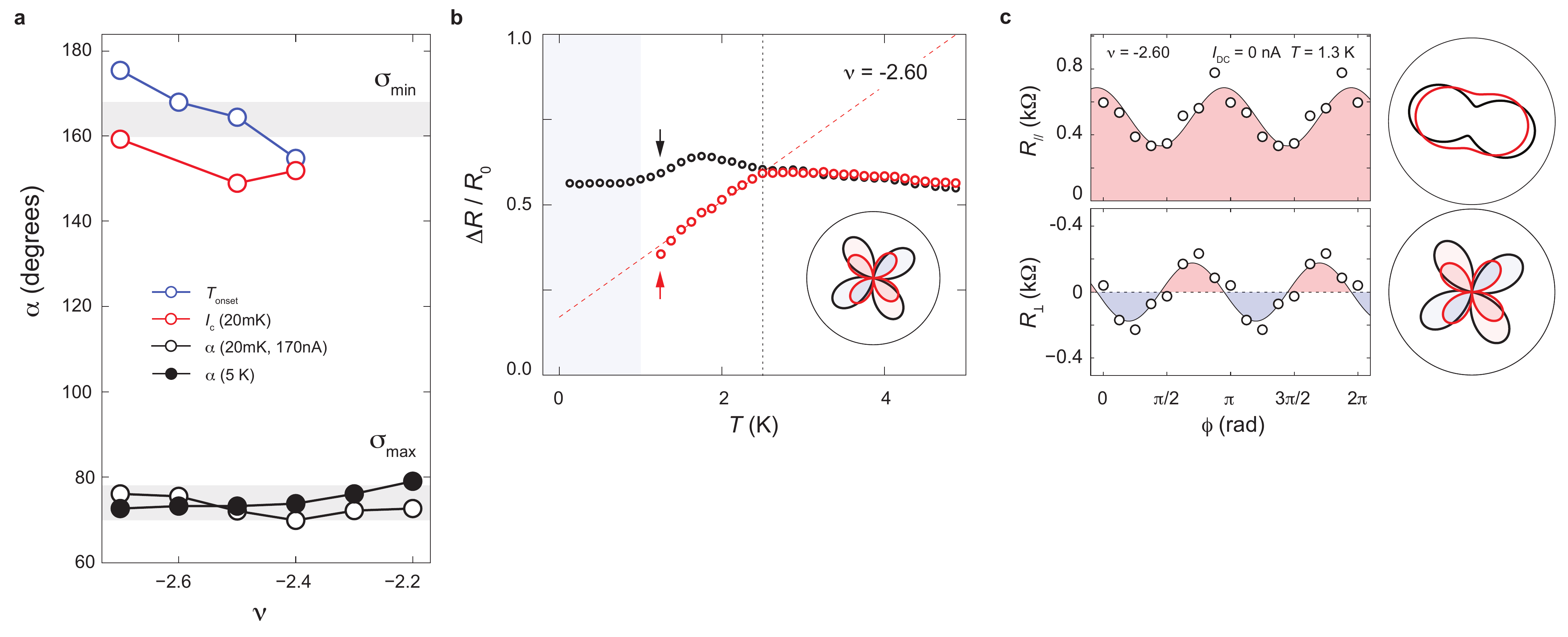}
\caption{\label{Dome}{\bf{Transport anisotropy across the superconducting regime.}}  $\alpha_I$, $\alpha_T$, and $\alpha_M$ as a function of $\nu$ across the superconducting dome. $\alpha_I$ and $\alpha_T$ are defined according to Fig.~\ref{fig2}, while $\alpha_M$ is extracted from the angular dependence of \Rpara\ and \Rperp\ measured at $T=20$ mK and $I_{dc} = 170$ nA (open black circles), and at $T=5$ K and $I_{dc} = 5$ nA (solid black circles). 
(b) The temperature dependence of \DRR\ measured in the high-doping regime at $\nu=-2.6$. Red circles, measured with a small d.c. current ($I_{dc} = 5$ nA), reflect the onset of superconductivity in the fluctuation regime. Black circles, measured with a large d.c. current ($I_{dc} = 170$ nA), reflect the temperature dependence of the metallic phase where superconductivity is fully suppressed. 
(c) Angular dependence of \Rpara\ and \Rperp\ measured with a small d.c. current bias in the fluctuation regime at $\nu = -2.6$. The angular dependence influenced by superconducting transport, shown as red solid lines in the polar-coordinate plots, is compared to that of the metallic phase (black solid lines). The angular oscillation in the \Rperp\ channel directly reflects the strength of anisotropy. This comparison indicates that the onset of superconductivity in the high-doping regime suppresses anisotropy compared to the metallic phase, consistent with the reported behavior in Fig.~\ref{fig2}. }
\end{figure*}

\begin{figure*}
\includegraphics[width=1\linewidth]{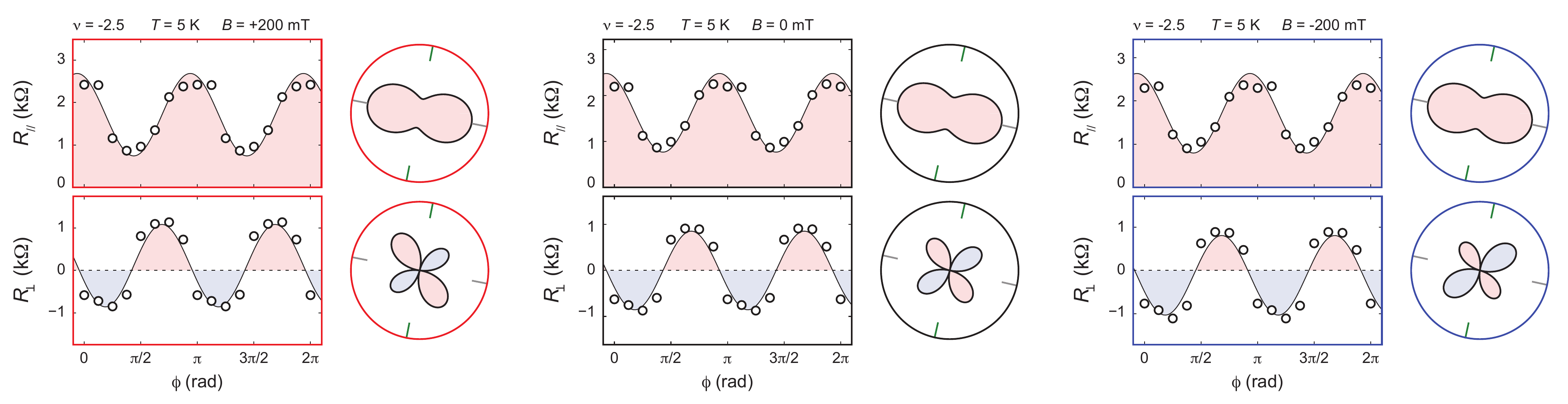}
\caption{\label{Bdependence}{\bf{Magnetic field and transport anisotropy.}} Angular dependence of \Rpara\  (top) and \Rperp (bottom) measured at $\nu=-2.5$ and $T=5$K. Left panels measured at $B=+200$mT, middle panels measured at $B=0$, right panels measured at $B=-200$mT. The angular dependence of \Rpara\  is insensitive to magnetic field. And as expected, the angular dependence of \Rperp\  acquires a shift in the averaged resistance when magnetic field is non-zero: positive for $B=+200$mT and negative for $B=-200$mT. }
\end{figure*}

\begin{figure*}
\includegraphics[width=0.67\linewidth]{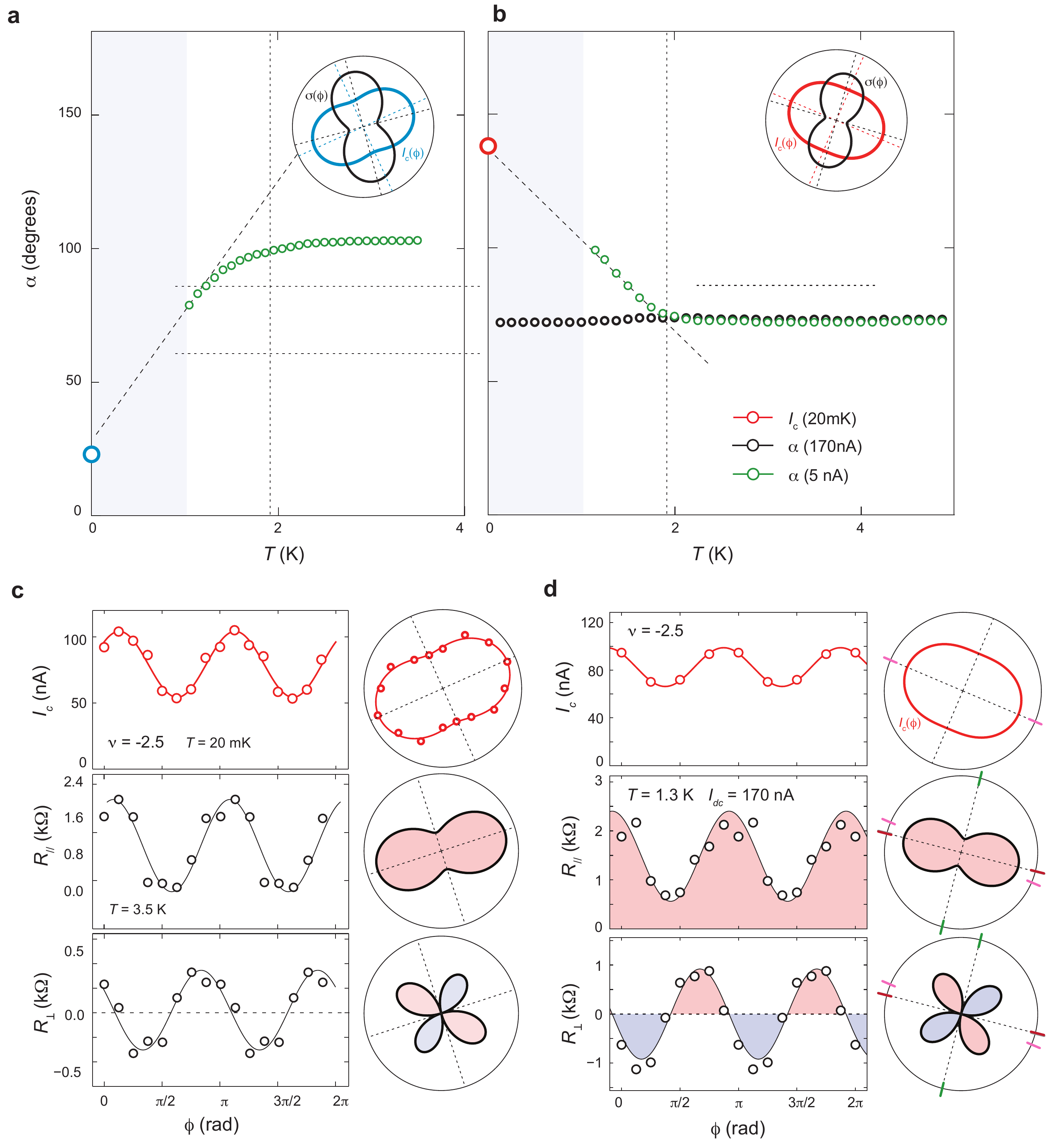}
\caption{\label{SChysteresis}{\bf{Hysteretic transition in the superconducting anisotropy.}} (a-b) Principal axis orientation $\alpha$ extracted from the angle-resolved transport, as a function of temperature, measured during first (a) and second (b) cool down. Green open circles denote $\alpha$ extracted from measurement with small a.c. current $I_{ac}=5$nA. Black open circles denote $\alpha$ extracted from measurement with superconductivity suppressed with large d.c. current $I_{dc}=170$nA. Blue and orange open circles denote $\alpha_I$ extracted from angle-resolved transport of critical current during first and second cool down. (c-d) Angular dependence of critical current $I_c$ (top) measured at $T=20$mK, \Rpara\ (middle) and \Rperp\ (bottom) measured at $T=3.5$K, during the first cool down (c) and second cool down (d). All data measured at the optimal doping of superconductivity, $\nu=-2.5$.}
\end{figure*}

Fig.~\ref{Dome}a illustrates the evolution of $\alpha_I$ and $\alpha_T$ as a function of band filling $\nu$ across the superconducting dome. The optimal doping of the superconducting phase, characterized by the maximum $T_c$ and $I_c$, is observed near $\nu = -2.5$. The angular dependence discussed in the main text is measured near this optimal doping. Here, we explore the evolution of superconducting anisotropy across the entire superconducting regime. For simplicity, we will refer to the band filling ranges of $\nu < -2.5$ and $\nu > -2.5$ as the high-doping and low-doping regimes, respectively.

In the high-doping regime, we observe the same angular interplay as at optimal doping  (Fig.~\ref{fig2}), where both $\alpha_I$ and $\alpha_T$ align with the direction of minimum conductivity in the metallic phase. This angular alignment is further supported by the behavior of transport anisotropy in the fluctuation regime. As illustrated in Fig.~\ref{Dome}b and d, the onset of superconducting transport in the fluctuation regime suppresses \DRR\ relative to its value in the metallic phase. These observations in the high-doping regime provide additional corroboration for the angular interplay observed at optimal doping.

In stark contrast, the superconducting transport in the low-doping regime exhibits distinct behavior. As illustrated in Fig.~\ref{Dome}a, $\alpha_I$ and $\alpha_T$ in the low-doping regime deviate from the direction of minimum conductivity in the metallic phase. Furthermore, in the fluctuation regime, \DRR\ measured at $\nu = -2.37$ appears to be enhanced by the influence of superconducting transport. 

The discrepancy between the low- and high-doping regimes could have a trivial explanation. Near half-filling of the moir\'e band, the carrier density resets, resulting in a low carrier density in the low-doping regime. This low carrier density makes the sample more susceptible to developing an inhomogeneous distribution, particularly at the onset of superconductivity. Such inhomogeneity introduces an extrinsic mechanism for superconducting anisotropy to deviate from the direction of minimum conductivity in the metallic phase.

\subsection{IV. Determination of the onset temperature}

Superconductivity can be suppressed either by applying a large d.c. current bias $I_{dc} = 170$ nA or by using an out-of-plane magnetic field of $B = 200$ mT. In the presence of a magnetic field, we employ methods described in Ref.~\cite{Vafek2023anisotropy} to extract the conductivity matrix, extending beyond the angular dependence described by Eq.~1 and 2. Importantly, the transport anisotropy obtained from measurements at $B = 200$ mT and $I_{dc} = 5$ nA is consistent with that observed at $B = 0$ and $I_{dc} = 170$ nA, demonstrating robustness in the observed angular dependence of the transport response.

Along the same vein, \Tonset\ can be determined by suppressing superconductivity either with a large d.c. current bias of $I_{dc} = 170$ nA or by applying an out-of-plane magnetic field of $B = 200$ mT. Notably, the angular dependence of \Tonset\ remains consistent regardless of the method used for analysis.

As shown in Fig.~\ref{fig2}e, the $R-T$ curves measured at $B = 200$ mT and $I_{dc} = 5$ nA is consistent with those measured at $B = 0$ and $I_{dc} = 170$ nA above \Tonset. However, these two curves bifurcates below \Tonset. This bifurcation could be reflective of a magnetic order at $B = 0$, which is suppressed by the presence of an out-of-plane $B$ field.

\subsection{V. Angular interplay between the zero-field diode effect and nematicity}

In tTGL near the magic angle, the correlated insulator near half-filling of the electron-doped moir\'e band is shown to exhibit orbital magnetism ~\cite{Zhang2024nonreciprocity}. This magnetic order leads to interesting transport responses in the nearby superconducting phase, as illustrated in Fig.~\ref{Diode}. 

Unlike the hole-doped superconductor investigated in the main text, predominantly exhibits reciprocal transport responses, the electron-doped superconductor exhibits notable nonreciprocity with an intriguing angular symmetry.  While the I-V curve is fully symmetric along $\phi = 45^{\circ}$, as shown in the top panel of Fig.~\ref{Diode}a,  the superconducting phase behaves like a forward diode at $\phi = 112.5^{\circ}$ (middle panel of Fig.~\ref{Diode}a), whereas an asymmetry characteristic of a reverse diode emerges at $\phi = 0^{\circ}$ (bottom panel of Fig.~\ref{Diode}a).

To quantify the asymmetry of the I-V curves, we introduce an onset current, $I_0$. Operationally, $I_0$ is defined as the point where $dV/dI$ reaches $20\%$ of the peak value in the I-V curve. In Fig.~\ref{Diode}a, vertical red and blue lines are used to indicate the values of $I_c$ and $I_0$, respectively. The angular dependence of $I_c$ and $I_0$ is shown in Fig.~\ref{Diode}b, revealing an intriguing angular correlation. Maximum nonreciprocity in superconducting transport, indicated by the largest asymmetry in $I_0$, occurs along the same direction associated with minimum $I_c$. Conversely, along the direction of maximum $I_c$, which defines the preferred direction of superconducting transport, the I-V curve is highly symmetric, indicating vanishing nonreciprocity. 

This angular dependence is quite intriguing. Since the zero-field superconducting diode effect is associated a time-reversal breaking field ~\cite{Yuan2021diodes,DiodeTheoryPaper,Daido2022SCDiode,He2022SCDiode}, the observed angular interplay enables several important observations. First, the angular dependence suggests a potential competition between superconductivity and the electronic order that breaks time-reversal symmetry. The emergence of this electronic order, aligned with the direction marked by the green solid line, simultaneously enhances nonreciprocity while suppressing $I_c$. Second, the one-fold angular symmetry by $I_0$ implies an intrinsic connection between rotational and time-reversal symmetry breaking ~\cite{Chichinadze2024nonlinearHall}. A plausible explanation for this angular symmetry is a type of momentum space instability, which has been explored by previous reports ~\cite{Zhang2024nonreciprocity,Chichinadze2024nonlinearHall,Dong2021momentum,Jung2015momentum,Huang2023momentum}.

\begin{figure*}
\includegraphics[width=0.9\linewidth]{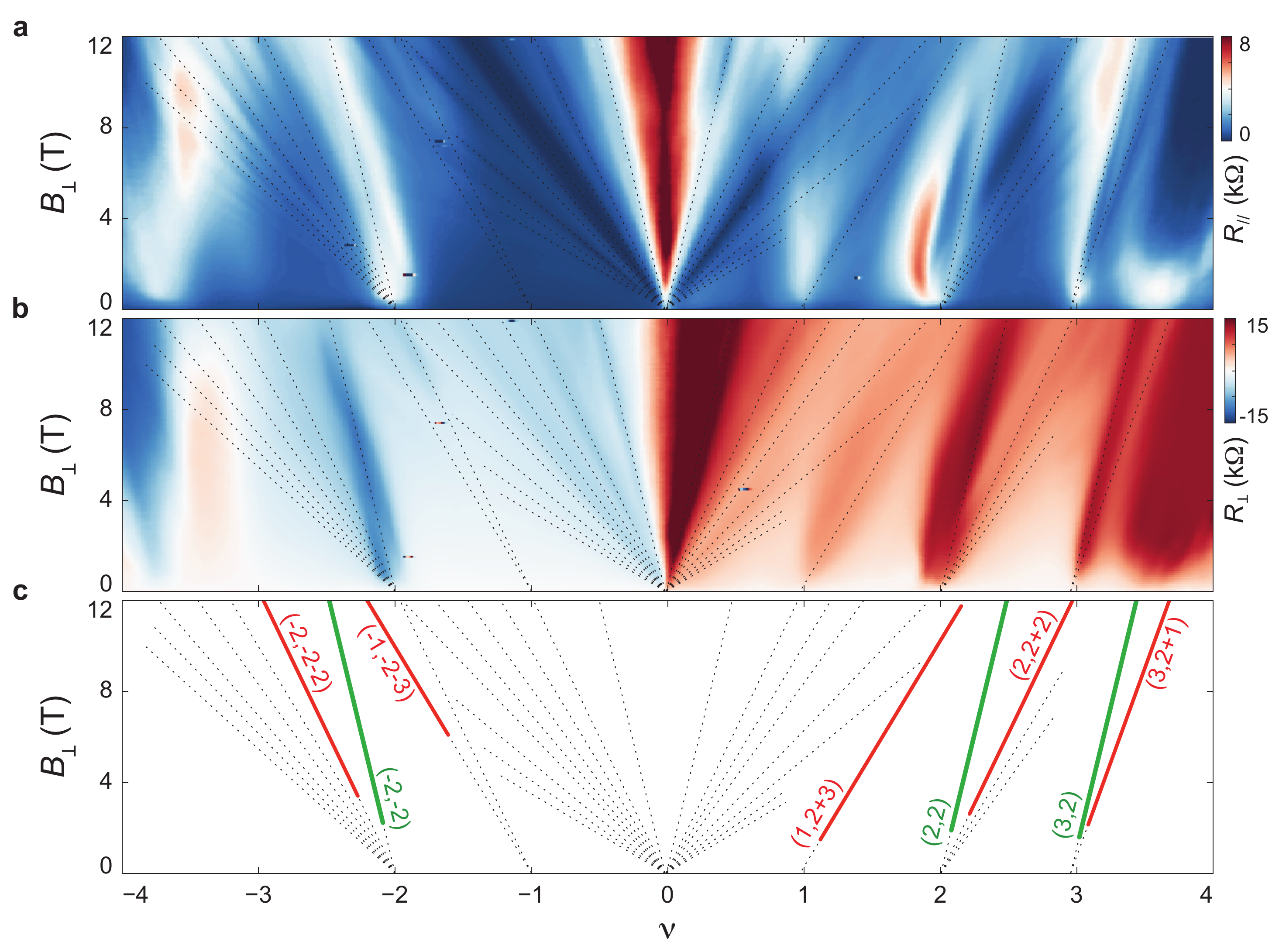}
\caption{\label{fan}{\bf{Isospin degeneracy across a mori\'e band.}} (a) $R_{\parallel}$ and (b) $R_{\perp}$ as a function of $\nu$ and $B_{\perp}$, measured on a sample with twist angle $\theta=1.45^o$ (one of the three samples shown in Fig.\ref{fig1}f). Incompressible states are manifested as minima in  $R_{\parallel}$ along with quantized plateaux in $R_{\perp}$. (c) Black dashed lines marks the trajectory of the most prominent incompressible states. Red and green solid lines are marked with a pair of quantum numbers ($t, s$) from the Diophantine equation $\nu = t \phi/\phi_{0} + s$, where $\nu$ is the filling factor $\nu_{tTLG}$ at the incompressible state ~\cite{Xie2021tblg,Spanton2018}. Green solid lines denote the trajectories of correlation-driven insulators, where a constant slope corresponds to a Chern number of 2, arising from the presence of a Dirac-like band.  Red solid lines indicate the most prominent Landau levels originating from each integer filling, with their slope corresponding to the degeneracy of the underlying Fermi sea.  Data measured at $T=20$mK and small a.c. curent bias of $5$nA along $\phi=0^\circ$. The presence of the Dirac-like band endows correlated insulators a Chern number of $2$. }
\end{figure*}

\begin{figure*}
\includegraphics[width=0.8\linewidth]{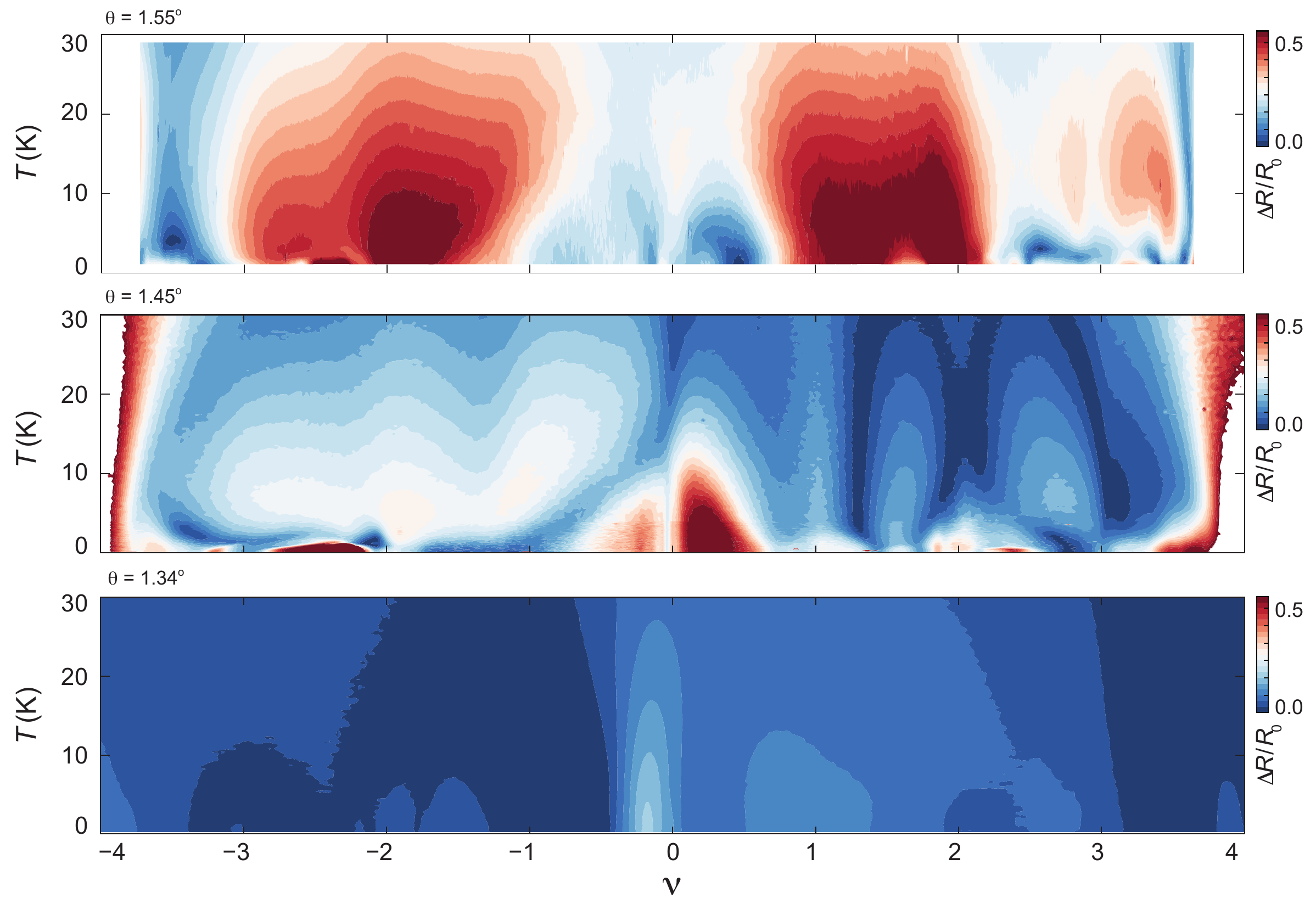}
\caption{\label{fignT}{\bf{Transport anisotropy across the phase space as the twist angle deviates from the magic angle.}} Transport anisotropy strength $\Delta R/ R_0$ as a function of $\nu$ and $T$ measured with three different samples with twist angle $\theta=1.55^\circ$ (top), $\theta=1.45^\circ$ (middle), and $\theta=1.34^\circ$ (bottom). }
\end{figure*}

\begin{figure*}
\includegraphics[width=0.95\linewidth]{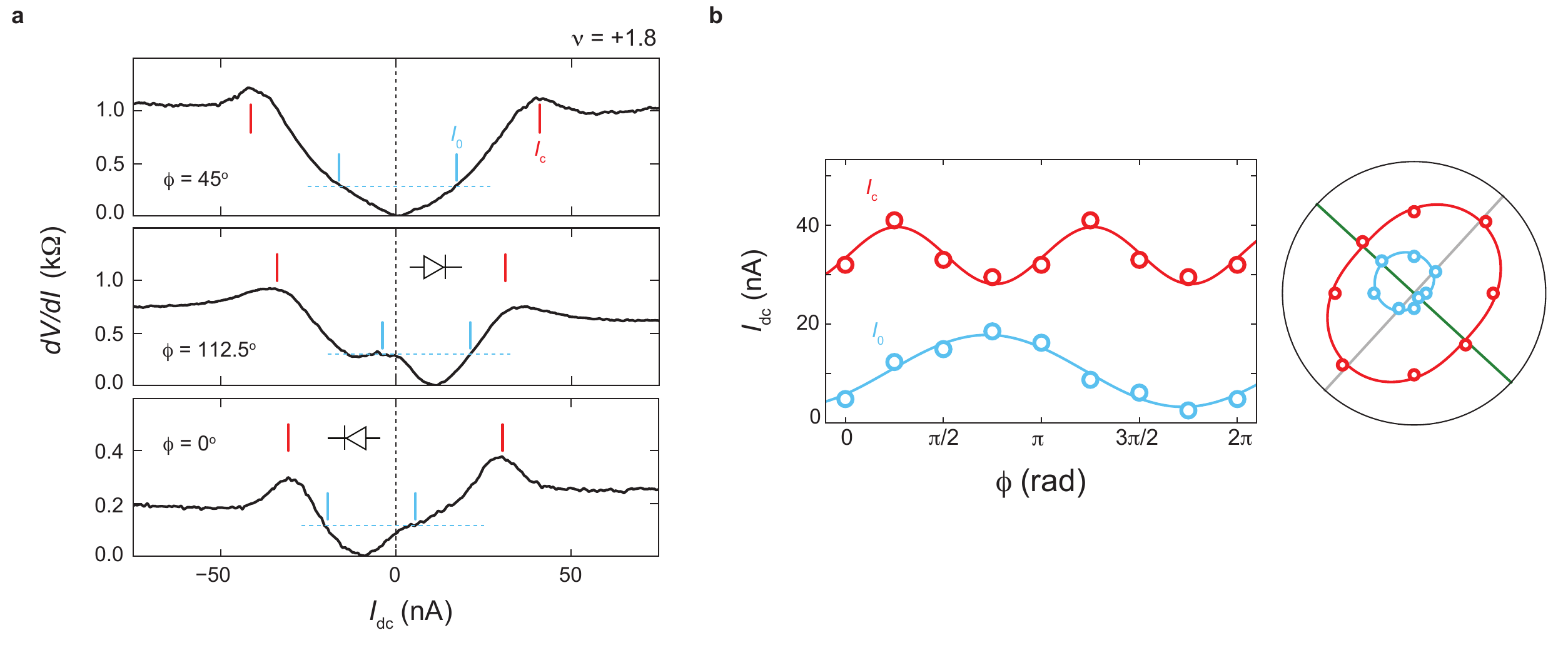}
\caption{\label{Diode}{\bf{Angular interplay between the zero-field diode effect and nematicity.}} (a) I-V curves measured at $\nu = 1.8$ along different current flow directions. The critical supercurrent $I_c$, marked by red vertical lines, is defined as the peak positions in the differential resistance $dV/dI$. To describe the asymmetry in the I-V curves, blue vertical lines mark $I_0$, which characterizes the onset of $dV/dI$ with increasing d.c. current. Operationally, $I_0$ is defined as the point where $dV/dI$ reaches $10\%$ of the peak value in the I-V curve. (b) The angular dependence of $I_c$ and $I_0$. Maximum $I_c$ defines the preferred direction of superconducting transport, along which the I-V curve is symmetric between forward and reverse current bias, resulting in zero nonreciprocity. Minimum $I_c$ defines the direction of suppressed superconducting transport, where the I-V curve exhibits maximum asymmetry, giving rise to the most prominent superconducting diode effect.  
}
\end{figure*}

\newpage

\newpage
\clearpage

\end{document}